\documentclass[11pt]{article}

\usepackage{amsfonts}
\usepackage[centertags]{amsmath}
\usepackage{amssymb}
\usepackage{cite}
\usepackage{cancel}
\usepackage{psfrag}
\usepackage{graphicx}
\usepackage{bm}
\usepackage{epsfig, multicol}
\usepackage[dvipsnames]{xcolor}
\usepackage{ulem}
\usepackage{hyperref}

\allowdisplaybreaks

\def\be{\begin{equation}}
\def\ee{\end{equation}}
\def\bea{\begin{eqnarray}}
\def\eea{\end{eqnarray}}
\def\ba{\begin{array}}
\def\ea{\end{array}}
\def\ben{\begin{enumerate}}
\def\een{\end{enumerate}}

\newcommand{\dsl}{\pa \kern-0.5em /} 
\usepackage{lipsum}
\usepackage{tikz}
\usetikzlibrary{arrows,decorations.pathmorphing,backgrounds,positioning,fit,petri,automata,shadows,calendar,mindmap,decorations.markings,calc}
\usepackage{enumerate}
\usepackage{comment}
\usepackage{lineno}
\newcommand{\rad}{R}
\newcommand{\zh}{z_{\rm h}}

\newcommand{\AdSrad}{L}
\newcommand{\lp}{\ell_{\rm P}}
\newcommand{\spar}{\lambda}
\newcommand{\x}{x}
\newcommand{\y}{y}

\newcommand{\w}{w}
\newcommand{\A}{P}
\newcommand{\ys}{u}
\newcommand{\qzh}{p}
\newcommand{\sparq}{\eta}

\numberwithin{equation}{section}

\title{Entanglement Entropy and Subregion Complexity in Thermal Perturbations around Pure-AdS Spacetime}%for Spherical Boundary Subregions in the AdS Black Hole Background}% Force line breaks with \\
\author{Aranya Bhattacharya${}^{\, a , \, b}$, Kevin T.~Grosvenor${}^{\, c}$, Shibaji Roy${}^{\, a ,\, b}$\\[.5truecm]
${}^a$Saha Institute of Nuclear Physics \\
1/AF Bidhannagar, Calcutta 700064, India \\[0.5cm]
{\rm and} \\[0.5cm]
${}^{\, b}$\textit{Homi Bhabha National Institute} \\
\textit{Training School Complex, Anushakti Nagar, Mumbai 400085, India} \\[0.5cm]
${}^{\, c}$Institut f\"ur Theoretische Physik und Astrophysik,\\
Julius-Maximilians-Universit\"at W\"urzburg, \\ Am Hubland, 97074 W\"urzburg, Germany\\[.5truecm]
{\sl E-mail:} aranya.bhattacharya@saha.ac.in, kevinqg1@gmail.com, \\ shibaji.roy@saha.ac.in}

\begin{document}

%%%%%%%%%%%%%%%%%%%%%%%%%%%%%%
%\begin{flushright}
%UM-TH-00-16\\
%SINP/TNP/00-20\\
%arXiv:YYMM.NNNN\\
%  
%\end{flushright}
%\preprint{}

\maketitle

\begin{abstract}
We compute the holographic entanglement entropy and subregion complexity of spherical boundary subregions in the uncharged and charged AdS black hole backgrounds, with the \textbf{change} in these quantities being defined with respect to the pure AdS result. This calculation is done perturbatively in the parameter $\frac{\rad}{\zh}$, where $\zh$ is the black hole horizon and $\rad$ is the radius of the entangling region. We provide analytic formulae for these quantities as functions of the boundary spacetime dimension $d$ including several orders higher than previously computed. We observe that the change in entanglement entropy has definite sign at each order and subregion complexity has a negative sign relative to entanglement entropy at each of those orders (except at first order or in three spacetime dimensions, where it vanishes identically). 

We combine pre-existing work on the ``complexity equals volume'' conjecture and the conjectured relationship between Fisher information and bulk entanglement to suggest a refinement of the so-called first law of entanglement thermodynamics by introducing a work term associated with complexity. This extends the previously proposed first law, which held to first order, to one which holds to second order. We note that the proposed relation does not hold to third order and speculate on the existence of additional information-theoretic quantities that may also play a role.
\end{abstract}

\tableofcontents

%%%%%%%%%%%%%%%%%%%%%%%%%%%%%%%%%%%%%%%%
% Introduction
%%%%%%%%%%%%%%%%%%%%%%%%%%%%%%%%%%%%%%%%

\section{Introduction}

The AdS/CFT correspondence \cite{Maldacena:1997re, Aharony:1999ti} has proven to be a powerful tool to study quantum field theories by passing to dual gravitational theories (e.g., \cite{Kovtun:2004de, Luzum:2008cw, Rangamani:2009xk}). The duality has also been used to address fundamental questions in quantum gravity (e.g., the black hole information paradox \cite{Barbon:2009zza, Almheiri:2012rt}) as well as aspects of quantum information theory \cite{nielsenbook}.

In this regard, a major breakthrough came from Ryu and Takayanagi \cite{Ryu:2006bv, Ryu:2006ef} who discovered a precise relationship between geometry, in the form of minimal-area surfaces, to entanglement entropy, a central concept in quantum information \cite{Bombelli:1986rw, Srednicki:1993im, Holzhey:1994we, Calabrese:2004eu, Eisert:2008ur, Nishioka:2009un, Takayanagi:2012kg, Witten:2018lha}. This relationship is given by
\begin{equation} \label{eq:HEE}
    S = \frac{A}{4 G_N} = \frac{2 \pi A}{\lp^{d-1}},
\end{equation}
where $G_N$ is the $(d+1)$-dimensional Newton's constant, $\lp$ is the Planck length and $A$ is the minimum area of a $(d-1)$-dimensional surface in the fixed time slice of AdS$_{d+1}$ which is homologous to and shares a common boundary with the subregion in the boundary CFT. This surface is referred to as the Ryu-Takayanagi (RT) surface and the right hand side of the equation is referred to as the holographic entanglement entropy (HEE). The RT prescription was fully justified and explained thereafter \cite{Lewkowycz:2013nqa}.

Corrections to the RT formula \eqref{eq:HEE} arising from bulk entanglement entropy of the RT surface were first proposed at leading order in the bulk Planck constant in \cite{Faulkner:2013ana}, extended to all orders in \cite{Engelhardt:2014gca} (see \cite{Jafferis:2015del} as well), and fully justified in \cite{Dong:2017xht}. These corrections have been checked in a number of cases \cite{Belin:2018juv}, although it is difficult to do this for general perturbations away from pure AdS due to complications in determining the modular Hamiltonian for general excited states. On the other hand, there has also been some work in finding higher-order corrections in the RT term itself (these are higher order in a small parameter measuring the perturbation away from pure AdS, e.g. the AdS black hole mass) \cite{Ghosh:2017ygi,He:2014lfa,Blanco:2013joa}. These corrections are also expected to be related to the change of energy density and pressure density of the gravity theory in the same way as the normal thermodynamic entropy is related to the change of energy and other thermodynamic variables.
  
In the business of calculating the HEE, typically two types of subsystems are considered, namely the infinite strip and the ball subsystem. These cases were studied in detail in the AdS black hold background to first order in the black hole mass \cite{Bhattacharya:2012mi}. This was followed by a detailed analysis up to second order in a small perturbation away from pure AdS (e.g., as a pure metric perturbation, as one produced by a bulk scalar, or as one produced by a boundary current)~\cite{Blanco:2013joa}. 
               
Some recent works endeavor to capture important physics with these second-order effects. For example, motivated by the ``complexity equals volume'' proposal \cite{Susskind:2014rva, Brown:2015bva, Brown:2015lvg}, Alishahiha proposed the volume enclosed by the RT surface as the gravitational dual to the boundary computational complexity of the state of a subregion in the boundary CFT \cite{Alishahiha:2015rta}. For this reason, this volume is called the holographic subregion complexity (HSC). More precisely, the HSC is defined to be
\begin{equation} \label{eq:Cdef}
    C = \frac{V}{8\pi \AdSrad G_{N}} = \frac{V}{\AdSrad \lp^{d-1}},
\end{equation}
where $\AdSrad$ is the AdS radius and $V$ is the volume enclosed by the RT surface. The leading-order change in HSC for a spherical subregion comes at second order, which has led to connections with fidelity susceptibility~\cite{Alishahiha:2015rta} and to Fisher information~\cite{Uhlmann:1975kt, Hayashi, Petz, Lashkari:2015hha, MIyaji:2015mia, Alishahiha:2017cuk, Banerjee:2017qti}. Other definitions of subregion complexity have also been proposed, e.g., kinematic space complexity and topological complexity \cite{Abt:2017pmf, Abt:2018ywl}.
  
Complexity is a notoriously difficult concept to define in quantum field theory in a way that does not appear to hinge on various arbitrary choices. Ordinarily, the measure of complexity involves minimizing the number of unitary transformations (within some choice of such transformations) required to transform the state of a system from some choice of reference state to the desired target state. In the context of the AdS/CFT correspondence, the cleanest aspect of this definition of complexity is the target state: we are clearly interested in CFT states which have known AdS duals. Our hope is that we can gain some insight into complexity by studying it perturbatively around holographic states. 

In a sense, what we have in mind here is a subregion-reduced version of the idea explored in \cite{Bernamonti:2019zyy}. This latter work tries to extract data about the so-called ``cost function'' \cite{nielsenbook, nielsen2005geometric, nielsen2006quantum}, which was introduced to describe ``minimal paths'' between reference and target states in state space, by studying the behavior of complexity under small variations in the target state. Keeping the reference state fixed, the variation in complexity is controlled just by the endpoint of the optimal path in state space. This result has been dubbed the ``first law of complexity'' \cite{Bernamonti:2019zyy}. More concretely, a set of coordinates $x^a$ is introduced on the space of unitary transformations $U(x^a)$ from some reference state $| \psi_{\rm R} \rangle$, which can also be interpreted as a set of coordinates on the space of states $U(x^a) | \psi_{\rm R} \rangle$. The path from the reference state $| \psi_{\rm R} \rangle$ to the target state $| \psi_{\rm T} \rangle = U_{\rm T} | \psi_{\rm R} \rangle$ minimizes the cost $\int_{0}^{1} ds \, F ( x^a (s) , \dot{x}^a (s) )$ of paths between the reference and target state, where $F$ is some ``cost function''. Under a small variation $\delta x^a$ of the target state, the leading-order change in complexity is 
\begin{equation}
    \delta C = p_a \delta x^a \bigr|_{s=1} \qquad\text{with} \qquad p_a = \frac{\partial F}{\partial \dot{x}^a}.
\end{equation}
Geometrically, then, $\delta C$ is related to the angle between the tangent to the optimal path at the target state and the displacement vector describing the variation away from the target. If $\delta C = 0$ at leading order, as is the case for the spherical subregion, then the tangent to the optimal path and the displacement of the target state are orthogonal at this order. The next-to-leading term is related to second derivatives of $F$ (for details, see \cite{Bernamonti:2019zyy}). This keeps going, of course, with higher order corrections to (subregion) complexity being related to higher derivatives of the cost function. Thus, we gain some insight into what the cost function might look like by studying small changes in holographic states. 

In \cite{Bernamonti:2019zyy} the starting target state was the ground state of the CFT, dual to pure AdS, and the perturbation was introduced by a bulk scalar field excitation corresponding to a coherent state. The complexity was that of the entire state, not a subregion-reduced state. Furthermore, that calculation was done within the ``complexity equals action'' framework. In our present work, we focus on thermal perturbations around pure AdS, our states will be reduced to a spherical boundary subregion, and we will be working within the ``complexity equals volume'' framework. Nevertheless, we share the same goal of studying the behavior of (subregion) complexity in the vicinity of holographic states to gain some insight into what paths from some reference state to a target state might look like. 

Consider the changes in HEE and HSC for a spherical subregion in the uncharged black hole background. The leading-order (LO) result for the HEE is what is often referred to as the ``first law of entanglement''~\cite{Blanco:2013joa}:
\begin{equation}
    \Delta \langle H \rangle = \Delta S \ \text{(at leading order),}    
\end{equation}
where $\langle H \rangle$ is the expectation value of the modular Hamiltonian in the state. At next-to-leading-order (NLO) for the HEE comes Fisher information, which has been related holographically to canonical energy~\cite{Lashkari:2015hha} and to bulk entanglement~\cite{Banerjee:2017qti}. For the HSC, the only thing known is that it vanishes identically for a spherical subregion in the AdS${}_3$ black hole background, the first-order term vanishes identically, and the second-order result is known only in the AdS${}_4$ black hole background~\cite{Alishahiha:2015rta, Banerjee:2017qti}. So far, the second-order result is not known in any other dimension and nothing is known at higher orders. We seek to fill in some of these gaps by computing second- and third-order corrections to the HSC in closed-form as functions of $d$. We provide closed-form formulae for the HEE up to third order as well as exact numerical expressions at fourth order. We also do these calculations for the case of a charged AdS black hole, which is an example in which the current perturbation also plays a role in addition to the metric perturbation. To the best of our knowledge, no such formulae have been heretofore reported in the literature.

For the uncharged black hole, we observe that $\Delta S$ is positive at first and third order and negative at second and fourth order. Of course, this is consistent with the first law of entanglement~\cite{Blanco:2013joa, OBannon:2016exv, Jaksland:2017nqx}. We expect that $\Delta S$ is positive at odd orders and negative at even orders. In contrast, $\Delta C$ is zero at first order, positive at second order and negative at third order. We expect that when $\Delta S \geq 0$, then $\Delta C \leq 0$, and vice versa. This finding is surprising from the point of view of the information-theoretic definitions of entanglement entropy and subregion complexity, which appear unrelated. Holographically, of course, the two appear related, at least superficially, since one is the area and the other the volume (assuming the ``complexity equals volume'' conjecture) of one and the same RT surface. However, a priori, the two quantities could have contained different and independent pieces of information about the RT surface. The switch in sign between $\Delta S$ and $\Delta C$ at each order indicates that they are not completely independent quantities; they are instead (anti-)correlated at least in their sign. This leads us to believe more that at each order, the change of HSC compensates the change of HEE from an information-theoretic point of view.

For the charged black hole, there are in-between orders which arise, so we can no longer speak of even and odd orders. Nevertheless, we still find that $\Delta S$ and $\Delta C$ appear at each order with opposite sign relative to each other.
 
We also consider the first law of entanglement thermodynamics proposed in~\cite{Allahbakhshi:2013rda} and which was shown therein to hold at first order. Combining previous work on holographic complexity~\cite{Alishahiha:2015rta} and and Fisher information~\cite{Banerjee:2017qti}, we propose a refinement of the first law of entanglement thermodynamics to include a general work term done on the system: $\Delta E = T \Delta S + W$.\footnote{This work term is different from the $V\Delta P$ term discussed in \cite{Allahbakhshi:2013rda}. This latter term appears at first order already and can be absorbed in the $\Delta E$ term by the equation of state with a suitable redefinition of the entanglement temperature.} These previous works naturally suggest that this work term be related to the change in HSC. However, now we find that the relation, which now holds at second order, does not hold at third order. This leads us to speculate that other information-theoretic quantities of interest might also play a role in a putative first law.

The remainder of the paper is organized as follows. In section 2, we discuss the computation of the embedding function of the RT surface. Detailed functional forms of the embedding function for spacetime dimensions 3 to 7 are given in the appendices. In section 3, we present our calculation of the change in HEE. In section 4, we present the corresponding results for HSC. The validity of the first law of entanglement and detailed entanglement thermodynamics is discussed in section 5, followed by our conclusions.

%%%%%%%%%%%%%%%%%%%%%%%%%%%%%%%%%%%%%%%%
% The Embedding Function
%%%%%%%%%%%%%%%%%%%%%%%%%%%%%%%%%%%%%%%%

\section{The Embedding Function}

First, we will discuss the case of the uncharged AdS${}_{d+1}$-Schwarzschild black hole (BH) of mass $m$ as a model example of a purely metric perturbation away from pure AdS (or, in the language of the dual field theory, a stress tensor perturbation). Then, we will discuss the case of a charged BH as well as a model example of a perturbation involving a current.\footnote{The reader may wonder why we do not simply consider a perturbation with a general stress tensor or current by relating the difference between the boundary metric in Fefferman-Graham coordinates from the flat Minkowski metric to the boundary stress tensor and current. This is done to leading and next-to-leading order for the stress tensor and to leading order for the current in \cite{Blanco:2013joa}. This requires writing the boundary metric in terms of products of stress tensors and currents at each order and determining the coefficients of these terms by matching to an explicit example, such as the AdS black hole. At higher orders, there can be many more terms than can be determined by one background. For example, at third order for a pure stress tensor perturbation, there are three possible terms: $T_{\mu\nu} T_{\alpha\beta} T^{\alpha\beta}$, $T_{\mu\alpha} T_{\nu\beta} T^{\alpha\beta}$ and $\eta_{\mu\nu} T_{\alpha\beta} T^{\beta\gamma} T_{\gamma}^{\alpha}$. The coefficients of these three terms in the expression for the boundary metric cannot be fully determined by the AdS black hole background since the latter only gives nonzero components for $\mu = \nu = 0$ and $\mu = \nu = i$. We encounter a similar problem for a mixed stress tensor and current perturbation, for example, in determining the coefficients of the three terms $T_{\mu\nu} J^2$, $T_{\alpha ( \mu} J_{\nu )} J^{\alpha}$ and $\eta_{\mu\nu} T_{\alpha\beta} J^{\alpha} J^{\beta}$ at order $3d-2$.} The latter is not a pure current perturbation, but is a mix of current and stress tensor perturbations. Generically, non-metric perturbations will inevitably be accompanied by metric perturbations via backreaction. Since we are interested in higher-order corrections to entanglement entropy and subregion complexity, we cannot in general ignore backreaction. Thankfully, the charged AdS black hole furnishes us with a fully backreacted solution involving a current perturbation.

We will expand the embedding function of the RT surface associated with a spherical boundary subregion of radius $\rad$ in the limit when $\rad$ is much smaller than the black hole radius of the background. For the uncharged black hole, this is equivalent to a ``small mass'' or ``low temperature'' expansion. For the charged case, the horizon radius depends on both the mass and charge of the black hole. However, the charge does not have to be small in our perturbative analysis, and so, even for the charged case, one may think of the perturbation as being in the smallness of the mass of the black hole. 

For the uncharged black hole, the first-order embedding function is known. We provide an analytic expression as a function of $d$ for the second-order result, which was not known prior to this work. While we have not been able to find a closed-form analytic expression for the third-order embedding function, we do supply explicit expressions for it in the cases of AdS${}_3$ to AdS${}_7$ to cover the cases of immediate import to AdS/CFT applications. 

For the charged black hole, there are orders which arise between the orders that are present in the uncharged case. What is called the $n$-th order in the uncharged case corresponds to what is more appropriately called the $nd$-th order in the charged case. It turns out that there are simple relationships between the $nd$-th order embedding functions in the charged case and the $n$-th order embedding functions in the uncharged case. The first in-between order is $2d-2$ with all others being sums of multiples of $d$ and of $2d-2$. Already at order $2d-2$, we are unable to give a closed-form analytic expressions for the corresponding embedding function. We will list out the embedding functions at order $2d-2$ and $3d-2$ for AdS${}_4$ to AdS${}_7$ in Appendix \ref{app:C}. 

\subsection{Uncharged AdS Black Hole}

We will work with the metric of a $(d+1)$-dimensional AdS-Schwarzschild black hole of mass $m$. The form of the metric is\footnote{A different metric is used in \cite{Bhattacharya:2012mi}, which is equivalent to this one up to first order in $m$ in the region $mz^d \ll 1$.}
\begin{equation} \label{eq:AdSBHmetric}
    ds^{2}= \frac{\AdSrad^{2}}{z^2}\left[ - f(z) \, dt^{2} + \frac{dz^{2}}{f(z)} + dr^{2}+ r^{2} d\Omega_{d-2}^{2}\right],
\end{equation}
where $\AdSrad$ is the AdS radius, $t \in (- \infty, \infty )$ is the time coordinate, $z \in (0, \zh )$ is the bulk radial coordinate with the boundary at $z=0$ and black hole horizon at $\zh$ given by $m \zh^{d} = 1$, $r$ is the boundary radial coordinate, $\Omega_{d-2}$ is the collection of boundary angular coordinates, and $f(z)$ is the blackening function
\begin{equation} \label{eq:f}
    f(z) = 1 - mz^d.
\end{equation}
We work with the entangling region $B$, which is a ball of radius $\rad$ (i.e., $0 \leq r \leq \rad$).
The corresponding RT surface is described by a spherically symmetric embedding function $z = z(r)$,
such that\footnote{The surface is often parametrized by $r = r(z)$ instead, which is well-adapted to the computation of counterterms \cite{Taylor:2016aoi} and the HSC \cite{Alishahiha:2015rta}. However, there is a technical issue in that the domain of $z$ itself receives corrections in $m$. As a consistency check, we have performed the second-order calculations using the $r(z)$ parametrization as well, yielding identical results. Higher-order computations were done purely in the $z(r)$ parametrization.}
\begin{align} \label{eq:bc}
    z( \rad ) = 0.
\end{align}
The area of the RT surface as a functional of $z(r)$ is\footnote{One drawback of the $z(r)$ parametrization is that it obscures the need for a cut-off at a small value $z = \epsilon$. Nevertheless, since we are computing only the difference relative to the pure AdS background, no such cut-off will be required.}
\begin{equation} \label{eq:areafunc}
    A = \Omega_{d-2} \AdSrad^{d-1} \int_{0}^{R} dr \frac{r^{d-2}}{z(r)^{d-1}} \sqrt{1 + \frac{z'(r)^2}{f(z(r))}},
\end{equation}
where $z' (r) = \frac{dz(r)}{dr}$ and $\Omega_{d-2} = \frac{2 \pi^{\frac{d-1}{2}}}{\Gamma \left( \frac{d-1}{2} \right)}$ is the volume of the $(d-2)$-sphere with unit radius. This area functional is extremized by solving the Euler-Lagrange equation. The embedding function is expanded as
\begin{equation}
    z (r) = z_0 (r) + \spar z_1 (r) + \spar^2 z_2 (r) + \spar^3 z_3 (r) + \cdots ,
\end{equation}
where the small expansion parameter is
\begin{equation}
    \spar = m \rad^d = \biggl( \frac{\rad}{\zh} \biggr)^d,
\end{equation}
and the Euler-Lagrange equation likewise expanded up to third order in $\spar$ to derive the equations satisfied by $z_0$, $z_1$, $z_2$ and $z_3$. It is convenient to measure lengths in units of $\rad$ and pass to the dimensionless variables
\begin{align}
    \x &\equiv \frac{r}{\rad}, &%
    \y ( \x ) &\equiv \frac{z(r)}{\rad}.
\end{align}
The boundary condition \eqref{eq:bc} becomes
\begin{align} \label{eq:bcy}
    \y ( 1 ) &= 0.
\end{align}
The function $\y_0 ( \x )$ is the pure AdS embedding
\begin{equation}
    \y_0 ( \x ) = \sqrt{1 - \x^2}.
\end{equation}
The equation for $\y_n ( \x )$ for $n \geq 1$ can be written as a Riemann-Papperitz equation \cite{siklos}
\begin{equation} \label{eq:RPE}
    \y_n'' + p( \x ) \y_n' + q( \x ) \y_n = \sigma_n ( \x ),
\end{equation}
where
\begin{subequations}
\begin{align}
    p( \x ) &= \frac{d-2-2 \x^2}{\x ( 1- \x^2 )}, \\
    q( \x ) &= - \frac{d-1}{( 1 - \x^2 )^2},
\end{align}
\end{subequations}
and $\sigma_n ( \x )$ is a driving function. 

The homogeneous part of this Riemann-Papperitz equation is identical for all orders, including the in-between orders that arise in the charged black hole case. This is the case because the homogeneous part of the equation for $\y_n$ comes from expanding just the pure-AdS part of the area functional to quadratic order in $\y_n$ and then taking the variation of the result with respect to $\y_n$. The genuinely difficult part of this equation is the driving function $\sigma_n ( \x )$ which depends on a complicated nested hierarchy of second-order differential operators acting on each previous term, each operator itself depending on even earlier terms. Needless to say, this problem increases in difficulty extremely quickly. The first-order embedding is relatively easy to solve in general $d$, taking on a rather simple closed form \eqref{eq:y1}. Already at second order the result \eqref{eq:y2} is fantastically more complicated. At third order, we are unable to find a closed form for the solution as a function of $d$. 

The problem is made simpler if we relax the requirement of finding a formula as a function of $d$ and instead compute the result for specific values of $d$. Of course, the computation increases in difficulty as $d$ increases, especially when $d$ is odd.\footnote{In fact, we are able to determine the third-order embedding for $d=12$, but not for $d=11$.} Nevertheless, we provide the third-order results for AdS${}_3$ to AdS${}_7$, thereby covering the cases most commonly considered in the context of applications of the AdS/CFT correspondence.\footnote{Of course, one could also perform numerics, which
would be a complementary approach (see e.g., \cite{Albash:2010mv}).
Our own endeavors in this direction, including attempts to repurpose the
shooting method code discussed in \cite{Ecker:2018jgh}, for example, were met
with technical difficulties. For instance, it seems extremely challenging
to get the numerical result for the first-order change in subregion
complexity to vanish identically, which makes the extraction of the
second-order change in subregion complexity very difficult. We are grateful
to Christian Ecker for his help and guidance with regard to the numerical
analysis.}

For $n=1$, the driving function is given by
\begin{equation}
    \sigma_1 ( \x ) = \frac{1}{2} ( 1 - \x^2 )^{\frac{d-3}{2}} \bigl[ 2(d-1) - (d+2) \x^2 \bigr].
\end{equation}
The first-order solution with boundary condition \eqref{eq:bcy} is
\begin{equation} \label{eq:y1}
    \y_1 ( \x ) = - \frac{(1 - \x^2 )^{\frac{d-1}{2}} ( 2 - \x^2 )}{2(d+1)}.
\end{equation}
The Fefferman-Graham version of this result is in \cite{Blanco:2013joa}. The $r(z)$ parametrization result is in \cite{Bhattacharya:2012mi,Alishahiha:2015rta}. We have verified that our $\y_1 ( \x )$ above is consistent with both of the aforementioned results. 

The second-order embedding function $\y_2 ( \x )$ does not contribute to the second-order change in HEE and is therefore not computed in \cite{Blanco:2013joa, Bhattacharya:2012mi}. It is needed for the second order change in HSC, which is studied in \cite{Alishahiha:2015rta}. This latter work gives the result without explicit computation for the second order change in HSC for $d=2$ and $d=3$ and the second order embedding function is not mentioned there either, presumably having been taken for granted.\footnote{In fact, we know that the change in HSC for $d=2$ should vanish identically at all orders since the BTZ black hole is locally equivalent to AdS${}_3$, the distinction being purely a topological one. Therefore, the relevant data point of genuine interest in \cite{Alishahiha:2015rta} is the change in HSC in $d=3$.} An explicit expression for the second-order embedding for AdS${}_4$, in the form $r_2 (z)$ is given in \cite{Banerjee:2017qti}. However, the result for the change in HSC therein is in conflict with that in \cite{Alishahiha:2015rta}. Therefore, we will give the expression for $\y_2 ( \x )$ for general dimension. To the best of our knowledge, this has not been done previously.

For $n=2$,
\begin{align}
    \sigma_2 ( \x ) = - ( 1 - \x^2 )^{d - \frac{7}{2}} \biggl( \frac{d^2 (d-1)}{(d+1)^2} &- \frac{5d^3 - 3d^2 + d -1}{2(d+1)^2} \x^2 \notag \\%
    &+ \frac{5d^3 + 6d^2 - 6d - 1}{4(d+1)^2} \x^4 - \frac{d-1}{4} \x^6 \biggr).
\end{align}
The solution is
\begin{align}
    \y_2 ( \x ) &= \frac{1}{16 \sqrt{\w}} \biggl( \frac{\sqrt{\pi}}{2^d} \frac{d(d-1)}{d+1} \frac{\Gamma ( d+1 )}{\Gamma \bigl( d + \frac{3}{2} \bigr)} \A + \frac{(d-1)(2d-1) (d-2)}{(d+1)^2} \A_0 \notag \\
    &\hspace{2cm} - \frac{3d^3 -15d^2 + 11d - 3}{(d+1)^2} \A_1 - \frac{2d^3 + 3d^2 -3d +2}{(d+1)^2} \A_2 - (d-1) \A_3 \biggr), \label{eq:y2} \\
    \intertext{where}
    \w &= 1- \x^2, \\
    \intertext{and}
    \A &= B \biggl( w; \frac{d}{2}, \frac{3-d}{2} \biggr), \\
    \A_n &= \A B \biggl( w; \frac{d}{2} -1+n, \frac{d-1}{2} \biggr) - \frac{2 w^{d-1+n}}{d(d-1+n)} {}_3 F_{2} \biggl( 1, \frac{3}{2} , d-1+n ; \frac{d}{2} + 1, d+n; w \biggr),
\end{align}
where $B(w;a,b)$ is the incomplete beta function. For convenience, we write out
these functions explicitly in Appendix \ref{app:A} for AdS${}_3$ to AdS${}_7$, which are the cases of greatest interest in the context of the AdS/CFT correspondence. We include the embedding
function up to second order in the $r(z)$ parametrization in Appendix \ref{app:A} as well. We have checked that the results for the embedding function at second order in both parametrizations are consistent with each other.

\begin{figure}[t]
\centering
\includegraphics[width=0.32\textwidth]{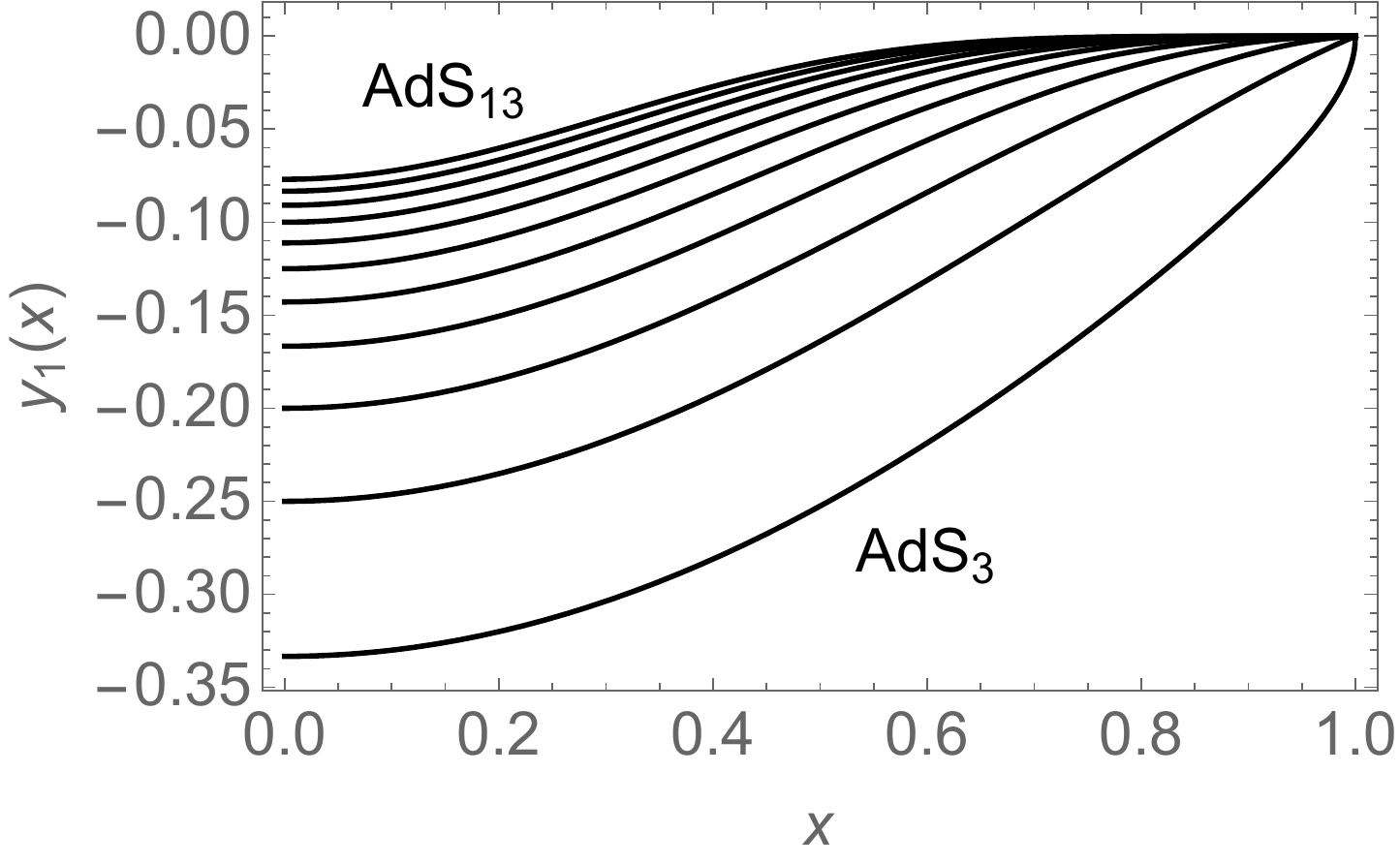}
\includegraphics[width=0.32\textwidth]{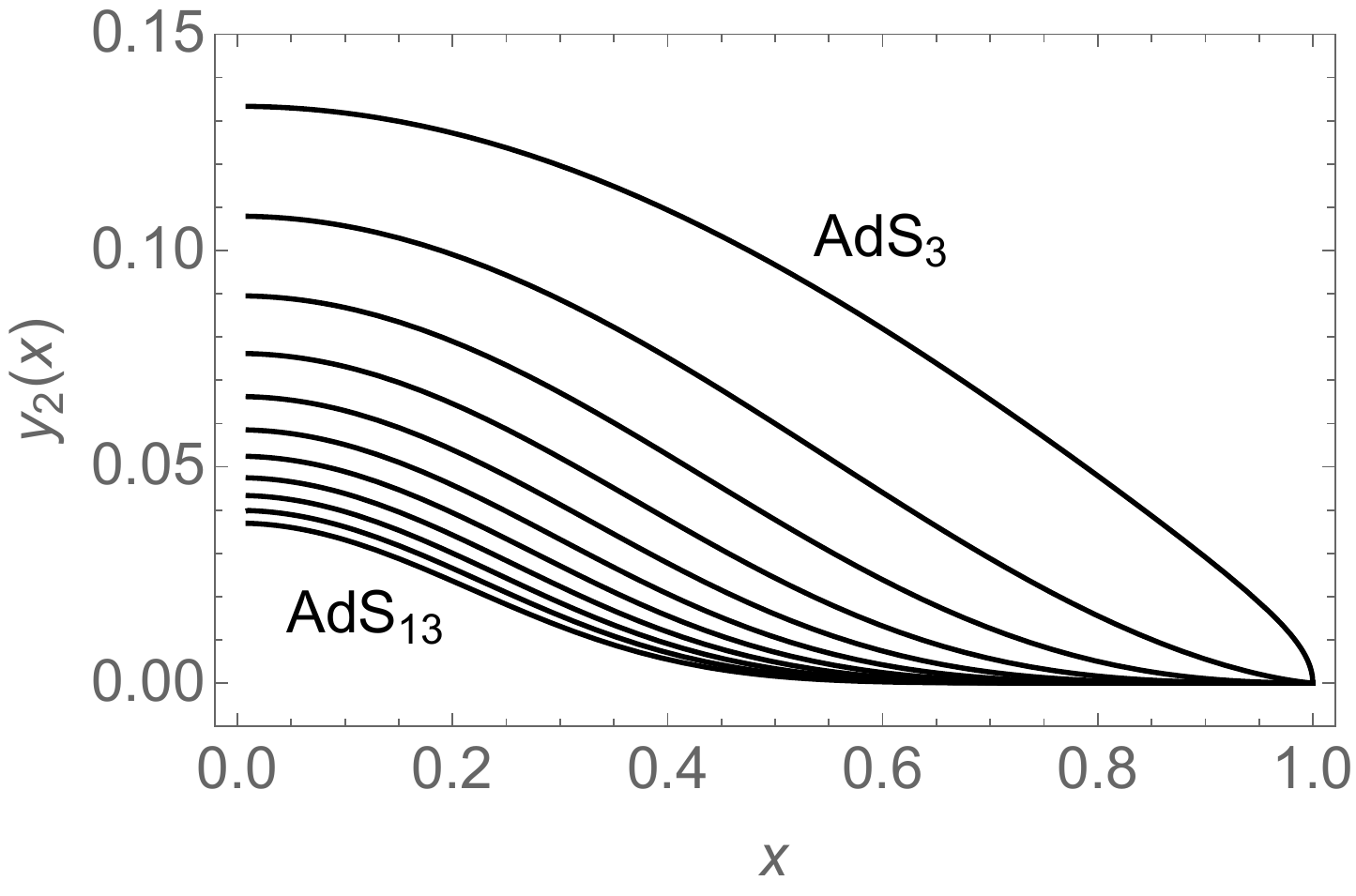}
\includegraphics[width=0.32\textwidth]{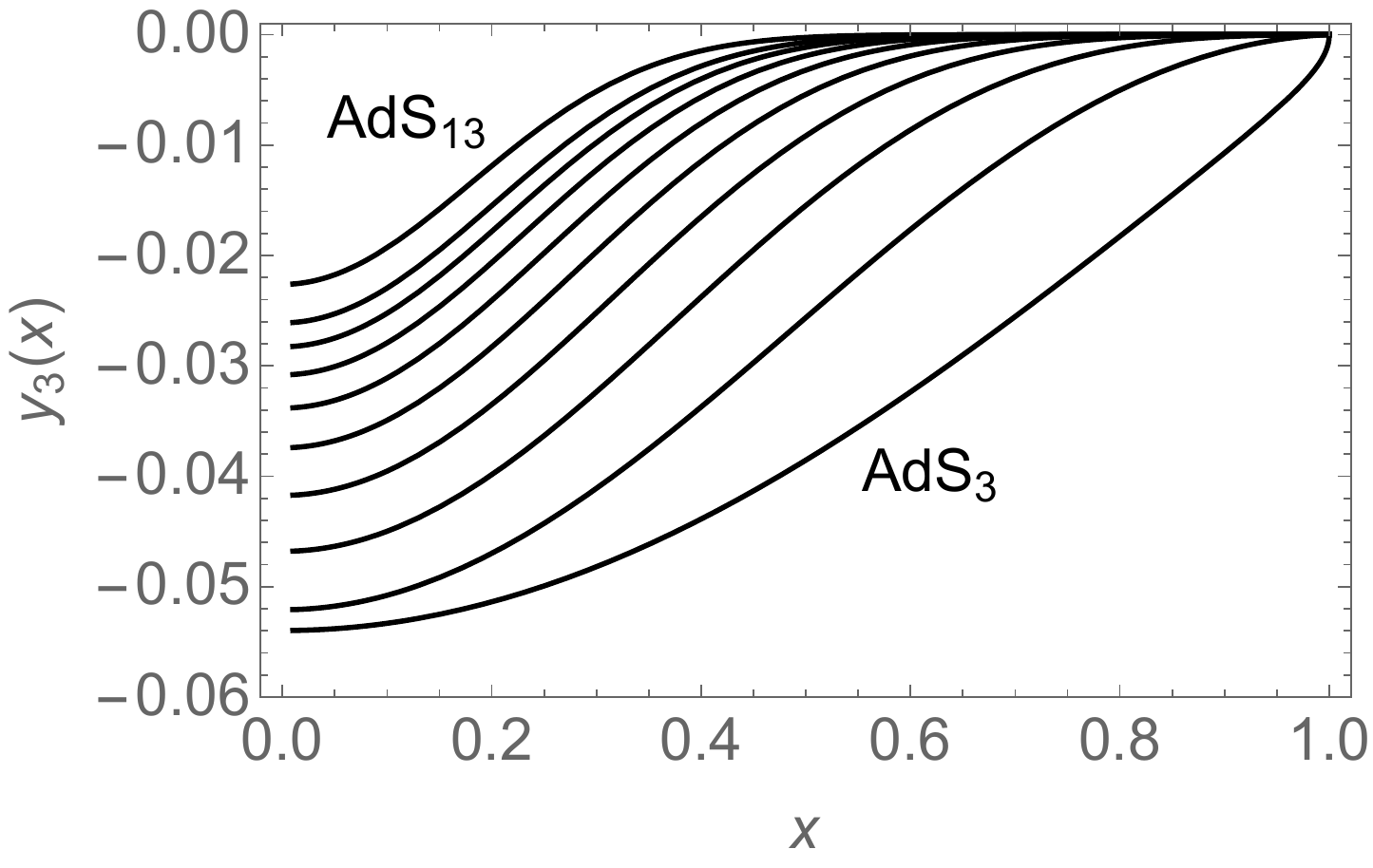}
\caption{Plots of $\y_1$, $\y_2$ and $\y_3$ in various dimensions. Note that we do not have $\y_3$ for AdS${}_{12}$.}
\label{fig:1}
\end{figure}

At third order, we do not have a general formula for the embedding function, but we give expressions for these in spacetime dimensions 3 to 7. Again, this covers all the usual cases of interest within the AdS/CFT context. At the third order, we do not provide the forms of the embedding functions in the $r(z)$ parametrization as we perform our calculations exclusively in the $z(r)$ parametrization, as in \cite{Blanco:2013joa}. To get a sense of the behavior of these higher-order embedding functions, we plot them in Figure \ref{fig:1}. Notice that the behavior of the embedding function near $\x =1$ for the case of AdS${}_3$ is very different compared to the higher-dimensional cases. This will turn out to be crucial in the analysis of boundary terms in subsection \ref{subsec:boundary}.

The general expression for $\y_2 ( \x )$ as a function of $d$ is very useful since we can use it to generate $\y_2$ for any value of $d$ without having to solve its defining differential equation each time. However, we are not actually able to perform the integrals needed to calculate the higher-order changes in HEE and HSC using the general form of $\y_2 ( \x )$. This complication will actually only be relevant to the second- and third-order changes in HSC. Therefore, we must infer formulae for these quantities from results at specific values of $d$.

\subsection{Charged AdS Black Hole}

We will now consider the charged AdS$_{d+1}$ black hole, which represents a class of perturbation away from pure AdS that also involves a boundary current in addition to a boundary stress tensor. The metric for the charged AdS BH takes on the same form as for the uncharged case \eqref{eq:AdSBHmetric} with the blackening function \eqref{eq:f} replaced with
\begin{equation} \label{eq:fq}
    f(z) = 1 - \bigl( 1+ q^{2} \zh^{2} \bigr) \frac{z^d}{\zh^{d}} +  q^{2} \zh^{2} \frac{z^{2d-2}}{\zh^{2d-2}}.
\end{equation}
Here the introduction of the new parameter $q$ is due to the presence of a current.\footnote{This term is related to the charge density carried by the horizon at $z= \zh$.} The gauge potential corresponding to this current has a single nonzero component,
\begin{equation}
      A_{0}(z) = \sqrt{\frac{2(d-1)}{d-2}} \, q \biggl( 1+ \frac{z^{d-2}}{\zh^{d-2}} \biggr).
\end{equation}
This is the same form of metric and gauge field used in \cite{Blanco:2013joa}. Note that there is no charged AdS${}_3$ black hole solution since the metric simply reduces to the uncharged case when $d=2$. For convenience, we define the dimensionless parameter 
\begin{equation}
    \qzh \equiv q \zh,
\end{equation}
which is treated as an $\mathcal{O}(1)$ constant.

Unlike the previous case (uncharged AdS BH), for which we defined $\spar = \frac{\rad^{d}}{\zh^{d}} = m \rad^{d}$, in this case, we define our dimensionless variables in the following way,
\begin{align}
    \x &\equiv \frac{r}{\rad}, &%
    \y ( \x ) &\equiv \frac{z(r)}{\rad}, &%
    \sparq &\equiv \frac{\rad}{\zh}.
\end{align}
Here, $\sparq$ is our perturbation parameter, which corresponds to the condition $\frac{\rad}{\zh} \ll 1$. Note that in the charged case, the orders in the expansion are controlled by two non-negative integers $n_1$, contributing $n_1 d$, and $n_2$, contributing $2n_2 (d-1)$. Thus, let us define the two-component vector $\Vec{n}$ and its ``size'' $| \Vec{n} |$ as the order at which it contributes:
\begin{align} \label{eq:nvec}
    \Vec{n} &= \binom{n_1}{n_2}, &%
    | \Vec{n} | &= n_1 d + 2n_2 (d-1).
\end{align}
Then, we expand the embedding function as
\begin{align} \label{eq:yqexp}
    \y ( \x ) &= \sum_{\Vec{n} = \Vec{0}}^{\Vec{\infty}} \sparq^{| \Vec{n} |} \y_{\Vec{n}} ( \x ) \notag \\
    &= \y_{(0,0)} ( \x ) + \sparq^d \y_{(1,0)} ( \x ) + \sparq^{2d-2} \y_{(0,1)} ( \x ) + \sparq^{2d} \y_{(2,0)} ( \x ) + \sparq^{3d-2} \y_{(1,1)} ( \x ) + \cdots,
\end{align}
with $\y_{(0,0)} ( \x )$ being the pure AdS embedding function,
\begin{equation}
    \y_{(0,0)} ( \x ) = \sqrt{1 - \x^2}.
\end{equation}
An important comparison between the expansion parameters of the uncharged and charged black holes is due here. As functions of their respective horizon radius $\zh$, one can see that $\spar = \sparq^{d}$. Thus, the orders which are integer multiples of $d$ in the charged case correspond to the orders present in the uncharged case and must reduce to the latter when $q=0$. On the other hand, the orders which are not simple integer multiples of $d$ (e.g., $(2d-2)$, $(3d-2)$, etc.) are not present in the uncharged AdS BH.

With this in mind, we solve the embedding in the same way as we did in the uncharged case. The orders at integer multiples of $d$ can be written simply in terms of the uncharged black hole embeddings in the following way,
\begin{equation}\label{eq:yyrel}
    \y_{(n,0)} ( \x ) = (1+ \qzh^{2})^{n} \y_{n}(\x),
\end{equation}
where $n$ is a non-negative integer.

For the newly appearing orders, however, it is difficult to come up with general expressions. Instead, we have to compute the embedding functions on a case by case basis for $d$ values starting from 3 to 6. We present the embedding functions $\y_{(0,1)} ( \x )$ and $\y_{(1,1)} ( \x )$ for the above mentioned $d$ values in Appendix \ref{app:C}. In this paper, we have considered up to order $(3d-2)$ for the charged black hole case. Again, to get a qualitative sense of the embedding functions, we plot different order embedding functions in Figure 2.

\begin{figure}[t]
\centering
\includegraphics[width=0.4\textwidth]{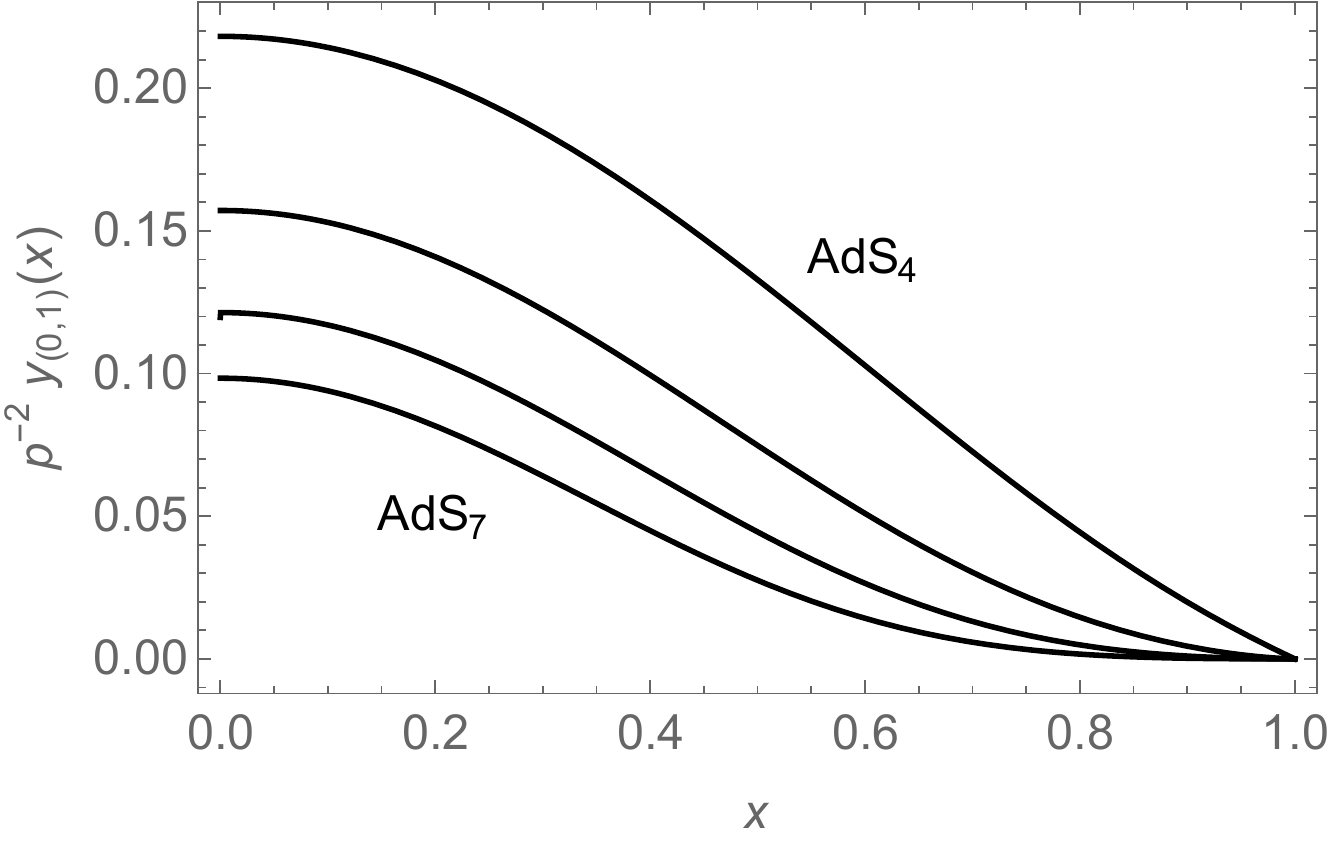} \qquad
\includegraphics[width=0.4\textwidth]{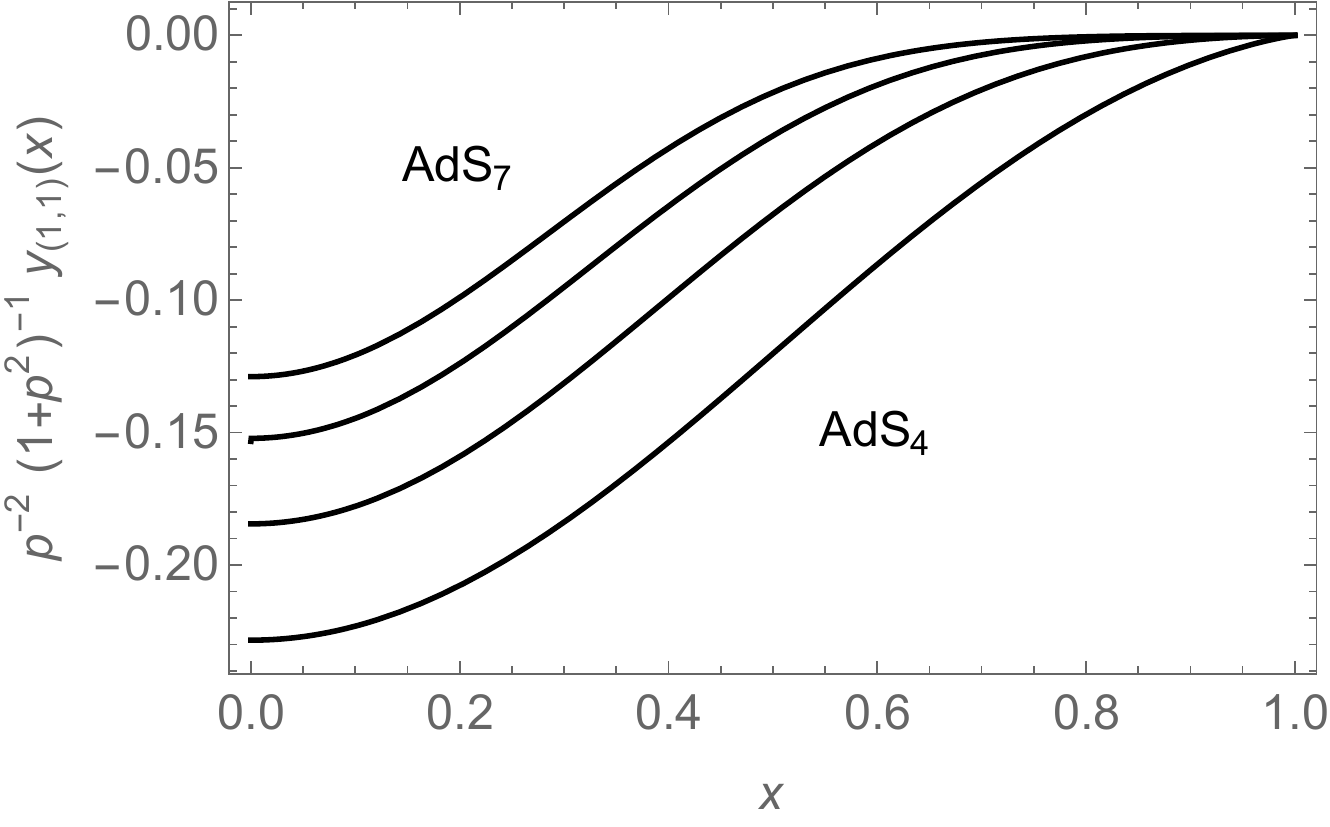}
\caption{Plots of $\frac{1}{\qzh^2} \y_{(0,1)}$ and $\frac{1}{\qzh^2 (1 + \qzh^2 )} \y_{(1,1)}$ in various dimensions.}
\label{fig:2}
\end{figure}

%%%%%%%%%%%%%%%%%%%%%%%%%%%%%%%%%%%%%%%%
% Holographic Entanglement Entropy
%%%%%%%%%%%%%%%%%%%%%%%%%%%%%%%%%%%%%%%%

\section{Holographic Entanglement Entropy}
\label{sec:HEE}
It is convenient to define the reduced HEE
\begin{equation}
    s \equiv \frac{S}{2 \pi \Omega_{d-2} \bigl( \frac{\AdSrad}{\lp} \bigr)^{d-1}}.
\end{equation}
In terms of the dimensionless variables,
\begin{equation} \label{eq:sgen}
    s = \int_{0}^{1} d \x \frac{\x^{d-2}}{\y ( \x )^{d-1}} \sqrt{1 + \frac{\y' ( \x )^2}{f( \y )}},
\end{equation}
where 
\begin{equation}
  f( \y )=1-\spar \y(\x)^{d}
\end{equation}
for the uncharged black hole and
\begin{equation}
    f( \y ) = 1 - (1 + \qzh^{2}) \sparq^{d} \y( \x )^{d} + \qzh^{2} \sparq^{2d-2} \y( \x )^{ 2d-2 }
\end{equation}
for the charged black hole.

%%%%%%%%%%%%%%%%%%%%%%%%%
% Uncharged AdS Black Hole
%%%%%%%%%%%%%%%%%%%%%%%%%

\subsection{Uncharged AdS Black Hole}

The explicit appearance of $\spar$ in \eqref{eq:sgen} is due to its appearance in the metric. When this factor of $\spar$ is expanded out, we refer to this as the ``metric contribution'' to the higher-order HEE. We introduce the notation $s_n$ to denote the metric contribution at order $\spar^n$. 

There is also the ``embedding contribution'', which comes from expanding the embedding function as
\begin{equation}
    \y ( \x ) = \y_0 ( \x ) + \spar \y_1 ( \x ) + \spar^2 \y_2 ( \x ) + \spar^3 \y_3 ( \x ) + \cdots.
\end{equation}
We pick out the term in $s_n$ of the form $\y_{n_1} ( \x ) \cdots \y_{n_k} ( \x )$ where $n_1 \leq \cdots \leq n_k$ and where some number of derivatives may act on the embedding functions. This term is denoted
\begin{equation}
    s_{n,n_1 \cdots n_k},
\end{equation}
and is a term in $s$ of order $\spar^{n+n_1 + \cdots + n_k}$. We make two exceptions in the above notation regarding $\y_0 ( \x )$. The indices $n_i$ are taken to be nonzero as long as at least one of them is nonzero. In other words, as far as the indices $n_i$ are concerned, we ignore factors of $\y_0 ( \x )$ as long as we are extracting a term that contains at least one higher-order correction to the embedding function. Otherwise, we write only one 0 after the comma in the subscript. For example, $s_{0,0}$ is the pure AdS result, while $s_{1,11}$ is a term in $s$ that is of order $\spar^3$ and consists of first expanding the metric to first order and then expanding the embedding function and picking out the terms that are quadratic in $\y_1 ( \x )$ and its derivatives. Since we are only interested in the difference from pure AdS, we define\footnote{Note that counterterms must be subtracted from the pure AdS result if one wants to calculate that term by itself (see \cite{Taylor:2016aoi}). This must also be done for the AdS black hole background separately. One considers differences in HEE partly in order to avoid these complications.}
\begin{equation}
    \Delta s = s - s_{0,0}.
\end{equation}
We expand this out in powers of $\spar$,
\begin{equation} \label{eq:dsexpand}
    \Delta s = \spar \Delta s^{(1)} + \spar^2 \Delta s^{(2)} + \spar^3 \Delta s^{(3)} + \cdots.
\end{equation}
As argued in \cite{Blanco:2013joa}, to calculate the first-order change in HEE, one needs only the zeroth-order embedding function. In fact,
\begin{equation}
    \Delta s^{(1)} = s_{1,0} = \frac{1}{2} \int_{0}^{1} d \x \, \x^{d-2} \frac{\y_0 \y_0'{}^2}{\sqrt{1+ \y_0'{}^2}} = \frac{1}{2} \int_{0}^{1} d \x \, \x^d = \frac{1}{2(d+1)}.
\end{equation}
The reason why $\y_1$ does not contribute to $\Delta s^{(1)}$ is because, after integration by parts, its contribution vanishes by the equation of motion for $\y_0$. However, a boundary term is obviously generated in the course of integrating by parts. In fact, this boundary term does not vanish and must instead be subtracted out in order that the variational principle for $\y_0$ be well-defined. Furthermore, in principle, there is an infinite hierarchy of such subtractions at higher and higher order. We will discuss these boundary terms for both the uncharged and charged cases in Section \ref{subsec:boundary}.

For the same reason as above, to compute the second order change in HEE, one needs the embedding only up to first order.\footnote{The nontrivial relationship between the depth of the RT surface in the bulk and the radius $\rad$ of the entangling region complicates the disentangling of second-order contributions when using the $r(z)$ parametrization. Nevertheless, using this method we get results consistent with the $z(r)$ parametrization if we use the full embedding function to second order.} In the course of our analysis of boundary terms in Section \ref{subsec:boundary}, we will derive the useful relations \eqref{eq:s0n} and \eqref{eq:s011rel}. Using these relations, we find
\begin{equation}
    \Delta s^{(2)} = s_{0,2} + s_{0,11} + s_{1,1} + s_{2,0} = \frac{1}{2} s_{1,1} + s_{2,0}.
\end{equation}
Using \eqref{eq:s011rel} to solve for $s_{0,11}$ in terms of $s_{1,1}$ is a substantial simplification since the latter is operationally much easier to compute than is the former. Nevertheless, we still verified this relation explicitly in this case. The final result is
\begin{equation}
    \Delta s^{(2)} = - \frac{\sqrt{\pi}}{2^{d+4}} \frac{(d-1) \Gamma ( d+1)}{(d+1) \Gamma \bigl( d + \frac{3}{2} \bigr)}.
\end{equation}
For the third-order change, one finds
\begin{align}
    \Delta s^{(3)} &= s_{0,3} + s_{0,12} + s_{1,2} + s_{0,111} + s_{1,11} + s_{2,1} + s_{3,0} \notag \\
    &= s_{0,111} + s_{1,11} + s_{2,1} + s_{3,0}.
\end{align}
Indeed, the central result of Section \ref{subsec:boundary} states that the embedding function up to first order is sufficient to compute $\Delta s$ up to third order. The final result is
\begin{equation}
    \Delta s^{(3)} = \frac{(9d^2 - 19d + 6)}{192 (d+1)^2} \frac{\Gamma (d+1) \Gamma \bigl( \frac{d+1}{2} \bigr)}{\Gamma \bigl( \frac{3(d+1)}{2} \bigr)}.
\end{equation}
With $\y_2$, we ought to be able to compute $\Delta s^{(4)}$. However, as previously stated, we are unable to evaluate the necessary integrals using the general form of $\y_2$ in \eqref{eq:y2}. Already at this point, the results for specific values of $d$ are sufficiently complicated that we are unable to infer a general formula as a function of $d$. We relegate the results that we have for $\Delta s^{(4)}$ to Appendix \ref{app:ct}. 

To summarize,
\begin{subequations} \label{eq:ds}
\begin{align}
    \Delta s^{(1)} &= \frac{1}{2(d+1)}, \label{eq:ds1} \\
    \Delta s^{(2)} &= - \frac{\sqrt{\pi}}{2^{d+4}} \frac{(d-1) \Gamma ( d+1)}{(d+1) \Gamma \bigl( d + \frac{3}{2} \bigr)}, \label{eq:ds2} \\
    \Delta s^{(3)} &= \frac{(9d^2 - 19d + 6)}{192 (d+1)^2} \frac{\Gamma (d+1) \Gamma \bigl( \frac{d+1}{2} \bigr)}{\Gamma \bigl( \frac{3(d+1)}{2} \bigr)}. \label{eq:ds3}
\end{align}
\end{subequations}
The first- and second-order terms agree with \cite{Blanco:2013joa}. The third-order term is a genuinely new result. Note that $\Delta s^{(2)} \leq 0$, as required by the first law of entanglement \cite{Blanco:2013joa}. Interestingly, $\Delta s^{(3)}$ is positive and the $\Delta s^{(4)}$ results in Appendix \ref{app:ct} are all negative. It appears that $\Delta s$ is positive in odd orders and negative in even orders.

%%%%%%%%%%%%%%%%%%%%%%%%%
% Charged AdS Black Hole
%%%%%%%%%%%%%%%%%%%%%%%%%

\subsection{Charged AdS Black Hole}

For the charged AdS BH, the perturbation parameter is $\sparq$. Thus, the change of entanglement entropy with respect to pure AdS can be written as the following expansion,
\begin{equation}
    \Delta s = \sparq^{d} \Delta s^{(1,0)} + \sparq^{2d-2}  \Delta s^{(0,1)} + \sparq^{2d}  \Delta s^{(2,0)} + \sparq^{3d-2}  \Delta s^{(1,1)}+ \cdots,
\end{equation}
where we extend the notation for the expansion of the embedding function introduced in \eqref{eq:yqexp} to the change in HEE: $\Delta s^{(\Vec{n})}$ is the term in $\Delta s$ of order $| \Vec{n}|$, where $| \Vec{n} |$ was defined in \eqref{eq:nvec}.

Our goal in this case is to compute the change in HEE for the charged BH up to order $(3d-2)$.
As shown in \cite{Blanco:2013joa}, to compute the HEE up to order $(2d-2)$, it is enough to take only $\y_{(0,0)}( \x )$. As in the uncharged case, this fact is actually just one of a hierarchy of such facts, which is the central result of our analysis of boundary terms in Section \ref{subsec:boundary}. For example, to compute the change in HEE to order $(3d-2)$, it is enough to use the embedding up to order $d$, which is $\y_{(1,0)}(\x)$. This is due to the fact that the contribution of $\y_{(0,1)}(\x)$ to $\Delta s^{(1,1)}$ vanishes by virtue of the Euler-Lagrange equation defining $\y_{(1,0)} ( \x )$.  

We have already mentioned how to get $\y_{(1,0)}(\x)$ from the uncharged BH results in \eqref{eq:yyrel}. Using this, we can  determine the change of HEE up to our desired order. The following are the results up to order $(3d-2)$,
\begin{subequations} \label{eq:dsch}
\begin{align}
    \Delta s^{(1,0)} &= \frac{(1+ \qzh^{2})}{2(d+1)} = (1+ \qzh^{2})\Delta s^{(1)}, \label{eq:dsch1}\\
    \Delta s^{(0,1)}&=- \qzh^{2} \frac{d-1}{2} \pi^{\frac{d+1}{2}} \frac{\Gamma \bigl( \frac{d}{2} \bigr)}{\Gamma \bigl( d + \frac{1}{2} \bigr)}, \label{eq:dsch2}\\ 
    \Delta s^{(2,0)} &= - (1+ \qzh^{2})^{2}\frac{\pi^{\frac{1}{2}}}{2^{d+4}} \frac{(d-1) \Gamma ( d+1)}{(d+1) \Gamma \bigl( d + \frac{3}{2} \bigr)} = (1+ \qzh^{2})^{2}\Delta s^{(2)}, \label{eq:dsch3} \\
    \Delta s^{(1,1)} &= \qzh^2 \left(1+ \qzh^2\right) \frac{(3 d-5)   \Gamma (d) \Gamma \left(\frac{d+1}{2}\right)}{8 (d+1) \Gamma \left(\frac{3 d}{2}+\frac{1}{2}\right)}. \label{eq:dsch4}
\end{align}
\end{subequations}
We observe that for the charged BH, $\Delta s^{(1,0)}$ and $\Delta s^{(1,1)}$ are positive definite whereas $\Delta s^{(0,1)}$ and $\Delta s^{(2,0)}$ are negative definite. Another fact that we can observe from \eqref{eq:dsch1} and \eqref{eq:dsch3} is the relation between changes of HEE for uncharged and charged black holes. This can be generalized in the following way,
\begin{equation}\label{eq:ssrelat}
    \Delta s^{(n,0)}  = (1+ \qzh^{2})^{n}\Delta s^{(n)},
\end{equation}
where $n$ is an integer. This is an expected observation analogous to \eqref{eq:yyrel}. These results will be important again after we compute the change of subregion complexity.

%%%%%%%%%%%%%%%%%%%%%%%%%
% Boundary Terms
%%%%%%%%%%%%%%%%%%%%%%%%%

\subsection{Boundary Terms}
\label{subsec:boundary}

As in \cite{Blanco:2013joa}, we implicitly subtract off some boundary terms in the change in HEE. This is justified as long as we take care to do this consistently. We give two cautionary examples which demonstrate that consistency requires certain integral boundary terms be subtracted out. It should then be clear how to formalize these examples into a proof of the central results of this subsection:
\begin{enumerate}
    \item Uncharged BH: $\Delta s^{(n)}$ is determined by the embedding function up to and including $\y_{\lfloor \frac{n}{2} \rfloor}$;
    
    \item Charged BH: $\Delta s^{( \Vec{n} )}$ is determined by the embedding function up to and including $\y_{\Vec{m}}$, where $\Vec{m}$ is the highest possible order such that $| \Vec{m} | \leq \frac{| \Vec{n} |}{2}$.
\end{enumerate}
The relationship \eqref{eq:yyrel} between the embedding function for the uncharged case and the charged case implies that the first point above is actually a special case of the second point. In other words, the second point reduces to the first when $q = 0$. 

As a generalization of the fact discussed earlier that $\y_1$ does not contribute to $\Delta s^{(1)}$, consider the contribution of $\y_n$ to $\Delta s^{(n)}$ for $n \geq 1$, which is just
\begin{equation}
    s_{0,n} = \int_{0}^{1} d \x \biggl( \frac{\delta s_0}{\delta \y} \biggr|_{0} \y_n + \frac{\delta s_0}{\delta \y'} \biggr|_{0} \y_n' \biggr),
\end{equation}
where the symbol $|_{0}$ means ``set $\y = \y_0$''. Integrating by parts and ignoring boundary terms gives
\begin{equation}
    s_{0,n} = \int_{0}^{1} d \x \biggl[ \frac{\delta s_0}{\delta \y } - \biggl( \frac{\delta s_0}{\delta \y'} \biggr)' \biggr] \biggr|_0 \y_n.
\end{equation}
The expression in the square brackets is precisely the Euler-Lagrange equation defining $\y_0$, which therefore vanishes when evaluated on $\y = \y_0$. Thus,
\begin{equation} \label{eq:s0n}
    n \geq 1: \qquad s_{0,n} = 0.
\end{equation}
Of course, a boundary term was ignored in the process, which is given by
\begin{equation}
    s_{0,n}^{\text{bdy}} = \biggl( \frac{\delta s_0}{\delta \y'} \biggr|_0 \y_n \biggr) \biggr|_{\x =1} - \biggl( \frac{\delta s_0}{\delta \y'} \biggr|_0 \y_n \biggr) \biggr|_{\x =0}.
\end{equation}
Note that $\frac{\delta s_0}{\delta y'}$ vanishes at $\x = 0$ for the full function $\y$ and not just $\y_0$. Since $\y_n$ is finite at $\x = 0$, the boundary contribution at $\x = 0$ vanishes. However, even though $\y_n ( 1) = 0$, the boundary contribution at $\x = 1$ does not necessarily vanish because $\frac{\delta s_0}{\delta \y'}$ contains a factor of $\y^{1-d}$, which diverges as $\x \rightarrow 1$. Indeed, for $n=1$, one sees from \eqref{eq:y1} that $\y_1$ is $\y_{0}^{d-1}$ multiplied by a function which is finite at $\x =1$. The resulting boundary term is
\begin{equation}
    s_{0,1}^{\text{bdy}} = \frac{1}{2(d+1)},
\end{equation}
which happens to be exactly equal to $\Delta s^{(1)} = s_{1,0}$. If one were to include $s_{0,1}^{\text{bdy}}$, then one would overestimate $\Delta s^{(1)}$ by a factor of 2. 

For $n = 2$, one can show that the behavior of $\y_2$ in \eqref{eq:y2} near $\x = 1$ is $(1- \x^2 )^{d - \frac{3}{2}}$, whereas $\y_{0}^{1-d} \sim ( 1 - \x^2 )^{\frac{1-d}{2}}$. Thus, the boundary function behaves like $( 1 - \x^2 )^{\frac{d}{2} -1}$ near $\x = 1$ and thus the boundary term vanishes identically except for $d=2$ or AdS${}_3$. Since we do not have $\y_3$ or higher in closed analytic form as a function of $d$, we cannot prove that this holds in general, but we have verified up to AdS${}_{13}$ (not included AdS${}_{12}$) that the boundary term also vanishes when $n = 3$ except for AdS${}_3$. As was hinted at earlier, that AdS${}_3$ is a special case can be seen quite clearly in the plots of the higher-order embedding functions in Figure \ref{fig:1}. The intuition here is that the boundary term $s_{0,n}$ arises because $\y_n$ is not ``flat enough'' at $\x = 1$. Evidently, $\y_1$ is never flat enough, regardless of the value of $d$. On the other hand, for $n \geq 2$, $\y_n$ is flat enough except for AdS${}_3$, which is not flat at all. Nevertheless, the boundary term must be subtracted out anyway.

It should be clear why the boundary term must be subtracted out of the final result, or simply ignored in the first place. If this is not done, then the variational principle used to determine $\y_0 ( \x )$ is not well-defined. As in the case of the Gibbons-Hawking-York boundary term in General Relativity, the appropriate boundary term must be added (or, indeed subtracted) in order to provide a well-defined and consistent variational principle.

For our second example highlighting the technicalities of boundary terms, consider the contribution of $\y_n$ to $\Delta s^{(n+1)}$ for $n \geq 1$. Firstly, let us discuss how one derives the Euler-Lagrange equation for $\y_1$. The leading term quadratic in $\y_1$ is $s_{0,11}$, which is of order 2. We add to $s_{0,11}$ all the terms which are of order 2 and linear in $\y_1$, namely $s_{1,1}$. Finally, we take a variation of the sum $s_{0,11} + s_{1,1}$ with respect to $\y_1$. Let us first write this sum out:
\begin{align}
    s_{0,11} + s_{1,1} &= \int_{0}^{1} d \x \biggl( \frac{\delta^2 s_0}{\delta \y^2} \biggr|_{0} \y_{1}^{2} + \frac{\delta^2 s_0}{\delta \y \delta \y'} \biggr|_{0} \y_1 \y_1' + \frac{\delta^2 s_0}{\delta y'^2} \biggr|_{0} \y_1'{}^{2} + \frac{\delta s_1}{\delta y} \biggr|_{0} \y_1 + \frac{\delta s_1}{\delta \y'} \biggr|_{0} \y_1' \biggr).
\end{align}
The variation with respect to $\y_1$ gives
\begin{equation}
    \frac{\delta ( s_{0,11} + s_{1,1} )}{\delta \y_1} = 2 \frac{\delta^2 s_0}{\delta y^2} \biggr|_{0} \y_1 - \biggl( \frac{\delta^2 s_0}{\delta \y \delta \y'} \biggr|_{0} \biggr)' \y_1 - 2 \biggl( \frac{\delta^2 s_0}{\delta \y'^2} \biggr|_{0} \y_1' \biggr)' + \frac{\delta s_1}{\delta y} \biggr|_{0} - \biggl( \frac{\delta s_1}{\delta \y'} \biggr|_{0} \biggr)'
\end{equation}
The vanishing of the above variation is the Euler-Lagrange equation for $\y_1$. Note that the homogeneous part of the equation comes from $s_{0,11}$. In general, the homogeneous part of the equation for $\y_n$ when $n \geq 1$ comes from $s_{0,nn}$. It is therefore not surprising that the homogeneous part of the Riemann-Papperitz equation \eqref{eq:RPE} defining $\y_n$ is the same for all $n \geq 1$.

Now, note that the contribution of $\y_n$ to $\Delta s^{(n+1)}$ for $n \geq 1$ comes from $s_{0,1n}$ and $s_{1,n}$. The first of these contains a relative factor of 2 when $n=1$ versus when $n \geq 2$:
\begin{subequations}
\begin{align}
    n &= 1: & %
    s_{0,11} &= \int_{0}^{1} d \x \biggl( \frac{\delta^2 s_0}{\delta \y^2} \biggr|_{0} \y_{1}^{2} + \frac{\delta^2 s_0}{\delta \y \delta \y'} \biggr|_{0} \y_1 \y_1' + \frac{\delta^2 s_0}{\delta y'^2} \biggr|_{0} \y_1'{}^{2} \biggr), \\
    n &\geq 2: & %
    s_{0,1n} &= \int_{0}^{1} d \x \biggl( 2 \frac{\delta^2 s_0}{\delta \y^2} \biggr|_{0} \y_1 \y_n + \frac{\delta^2 s_0}{\delta \y \delta \y'} \biggr|_{0} ( \y_1 \y_n' + \y_n \y_1' ) + 2 \frac{\delta^2 s_0}{\delta y'^2} \biggr|_{0} \y_1' \y_n' \biggr). \label{eq:s01n}
\end{align}
\end{subequations}
The other contribution, $s_{1,n}$, is given for $n \geq 1$ by
\begin{equation}
    n \geq 1: \quad
    s_{1,n} = \int_{0}^{1} d \x \biggl( \frac{\delta s_1}{\delta y} \biggr|_{0} \y_n + \frac{\delta s_1}{\delta \y'} \biggr|_{0} \y_n' \biggr).
\end{equation}
Integrating by parts and ignoring boundary terms gives
\begin{subequations}
\begin{align}
    n &= 1: &%
    s_{0,11} &= \int_{0}^{1} d \x \biggl[ \frac{\delta^2 s_0}{\delta y^2} \biggr|_{0} \y_1 - \frac{1}{2} \biggl( \frac{\delta^2 s_0}{\delta \y \delta \y'} \biggr|_{0} \biggr)' \y_1 - \biggl( \frac{\delta^2 s_0}{\delta \y'^2} \biggr|_{0} \y_1' \biggr)' \biggr] \y_1, \\
    n &\geq 2: &%
    s_{0,1n} &= \int_{0}^{1} d \x \biggl[ 2 \frac{\delta^2 s_0}{\delta y^2} \biggr|_{0} \y_1 - \biggl( \frac{\delta^2 s_0}{\delta \y \delta \y'} \biggr|_{0} \biggr)' \y_1 - 2 \biggl( \frac{\delta^2 s_0}{\delta \y'^2} \biggr|_{0} \y_1' \biggr)' \biggr] \y_n, \label{eq:s01nIBP} \\
    n &\geq 1: &%
    s_{1,n} &= \int_{0}^{1} d \x \biggl[ \frac{\delta s_1}{\delta \y} \biggr|_{0} - \biggl( \frac{\delta s_1}{\delta \y'} \biggr|_{0} \biggr)' \biggr] \y_n.
\end{align}
\end{subequations}
When $s_{0,1n}$ and $s_{1,n}$ are summed together, then, for $n \geq 2$, the expression multiplying $y_n$ is precisely the Euler-Lagrange equation defining $\y_1 ( \x )$ and therefore the sum vanishes. For $n=1$ one has to multiply $s_{0,11}$ by 2 to get the same result. Therefore,
\begin{subequations} \label{eq:s01nrelation}
\begin{align}
    n &= 1: &%
    2s_{0,11} + s_{1,1} &= 0, \label{eq:s011rel} \\
    n &\geq 2: &%
    s_{0,1n} + s_{1,n} &= 0.
\end{align}
\end{subequations}
Of course, boundary terms were ignored to get the above result. These boundary terms should actually appear on the right hand side of the above equations, instead of 0. Nevertheless, these boundary terms have to be subtracted out anyway to yield a well-defined and consistent variational principle for $\y_1 ( \x )$.\footnote{Again, at least in the case of $n=2$ and $n=3$, which is as far as we have expanded the embedding function in this work, it turns out that the boundary terms that have been ignored above actually vanish identically except for AdS${}_3$. So, the process of subtracting out these boundary terms is only nontrivial for the case of AdS${}_3$.} In fact, we have an even more immediate sign that these boundary terms must be subtracted out: if not, then the result for $\Delta s^{(2)}$ for the AdS${}_3$ black hole would be $\frac{1}{120}$ instead of $- \frac{1}{180}$. This is a positive number, which violates the first law of entanglement stated in \cite{Blanco:2013joa}.

This argument generalizes completely to the following statement: the contribution of $\y_n$ to $\Delta s^{(n+m)}$ for $n \geq 1$ and $m < n$ vanishes. Also, for $m =n$, the contribution is just equal to $- s_{0,nn}$. The procedure is exactly the same as with $m=1$. The desired contribution is 
\begin{equation}
    \sum_{q=0}^{m} \sum_{P(m-q)} s_{q,P(m-q)n},
\end{equation}
where $P(m-q)$ stands for all partitions of $m-q$ into a list of integers which are non-decreasing read left to right. Meanwhile, the Euler-Lagrange equation for $\y_m$ is derived by taking the variation with respect to $\y_m$ of the exact same sum, but with $\y_n$ replaced with $\y_m$. The same analysis as for $m=1$ shows that
\begin{equation}
    \sum_{q=0}^{m} \sum_{P(m-q)} s_{q,P(m-q)n} = 
\begin{cases}
    0, & n \geq 1\ \text{and}\ m<n, \\
    -s_{0,nn}, & n \geq 1\ \text{and}\ m=n.
\end{cases}
\end{equation}
The sum does not simplify in general for $m > n$ and is generally nonzero.

This argument generalizes with only cosmetic changes to the charged case: the contribution of $\y_{\Vec{n}}$ to $\Delta s^{( \Vec{n} + \Vec{m} )}$ vanishes when $| \Vec{m} | < | \Vec{n} |$ and is equal to $- s_{0, \Vec{n} \Vec{n}}$ when $| \Vec{m} | = | \Vec{n} |$.

These statements are equivalent to central result stated at the beginning of this subsection.

%%%%%%%%%%%%%%%%%%%%%%%%%%%%%%%%%%%%%%%%
% Holographic Subregion Complexity
%%%%%%%%%%%%%%%%%%%%%%%%%%%%%%%%%%%%%%%%

\section{Holographic Subregion Complexity}

We now compute the change in HSC. The volume is given by
\begin{equation}
    V = \Omega_{d-2} \AdSrad^d \int_{0}^{\rad} dr \, r^{d-2} \int_{\epsilon \rad}^{z (r)} \frac{dz}{z^d \sqrt{f(z)}},
\end{equation}
where we have introduced a cut-off $\epsilon \rad$ near $z=0$. The HSC is related to this by \eqref{eq:Cdef}. We define the reduced HSC $c$ as the HSC measured in units of $\frac{\Omega_{d-2}}{d-1} \bigl( \frac{\AdSrad}{\lp} \bigr)^{d-1}$:
\begin{equation}
    c \equiv \frac{C}{\frac{\Omega_{d-2}}{d-1} \bigl( \frac{\AdSrad}{\lp} \bigr)^{d-1}}.
\end{equation}
In terms of the dimensionless variables,
\begin{equation}
    c = (d-1) \int_{0}^{1} d \x \, \x^{d-2} \int_{\epsilon}^{\y ( \x )} \frac{d \y}{\y^d \sqrt{f(y)}}.
\end{equation}
The subscript notation we defined for $s$ carries through for $c$. The blackening functions $f( \y )$ for uncharged and charged black hoels are as mentioned in Section \ref{sec:HEE}. An important point to remember here is that, in contrast to the HEE case, to calculate the HSC to some order, we require the embedding function up to that same order. The simplifications that arose in the HEE case were due to the fact that the embedding function is derived by minimizing the area integral. No such simplification will occur in general for the volume integral. Now we jump into specific results for the uncharged and charged BH in the following subsections. 

%%%%%%%%%%%%%%%%%%%%%%%%%%%%%%%%%%%%%%%%
% Uncharged Holographic Subregion Complexity
%%%%%%%%%%%%%%%%%%%%%%%%%%%%%%%%%%%%%%%%

\subsection{Uncharged AdS Black Hole}

The quantity of interest here is the change in going from the pure AdS case to the uncharged AdS black hole case,
\begin{equation}\label{eq:deltac}
    \Delta c = c - c_{0,0}.
\end{equation}
which is finite as $\epsilon \rightarrow 0$ and is at least first order in $\spar$:
\begin{equation}
    \Delta c = \spar \Delta c^{(1)} + \spar^2 \Delta c^{(2)} + \spar^3 \Delta c^{(3)} + \cdots .
\end{equation}
In fact, we find that the first-order term vanishes. This result was stated in \cite{Alishahiha:2015rta, Ben-Ami:2016qex} and demonstrated explicitly in \cite{Banerjee:2017qti}. Therefore, the change in HSC is at least second order. Again, since we are unable to compute the requisite integrals using the general formula for $\y_2$ in \eqref{eq:y2}, nor do we have a general formula for $\y_3$, we must infer the general formulae for $\Delta c^{(2)}$ and $\Delta c^{(3)}$ from the results at specific values of $d$. This might seem rather hopeless at first. However, we do have some amount of guidance from the pieces in $\Delta c^{(2)}$ and $\Delta c^{(3)}$ that depend only on $\y_0$ and $\y_1$, which we can compute exactly. This guidance is enough for us to determine the formulae in general. We use the results for AdS${}_3$ to AdS${}_7$, the cases of greatest interest in the AdS/CFT context, to come up with general formulae as functions of $d$. We then test these formulae in the cases of AdS${}_8$ to AdS${}_{13}$, excluding AdS${}_{12}$ for $\Delta c^{(3)}$ since $\y_3$ for AdS${}_{12}$ is too lengthy and complicated to compute the requisite integrals.\footnote{We are greatly indebted to Charles Melby-Thompson for his help in determining the form of $\Delta c^{(2)}$ for AdS${}_3$ to AdS${}_7$ before we had an expression for $\y_2$ in general $d$.} The results as functions of $d$ are
\begin{subequations} \label{eq:dc}
\begin{align}
    \Delta c^{(1)} &= 0, \\
    \Delta c^{(2)} &= \frac{\sqrt{\pi}}{2^{d+2} (d+1)} \frac{\Gamma \bigl( \frac{d+1}{2} \bigr)}{\Gamma \bigl( \frac{d}{2} - 1 \bigr)}, \label{eq:dc2} \\
    \Delta c^{(3)} &= - \frac{d(9d-4)(2d-3)(d-1)(d-2)}{192 (d+1)^2} \frac{\Gamma \bigl( d- \frac{3}{2} \bigr) \Gamma \bigl( \frac{d+1}{2} \bigr)}{\Gamma \bigl( \frac{3d}{2} +1 \bigr)}. \label{eq:dc3}
\end{align}
\end{subequations}
The AdS${}_3$ and AdS${}_4$ results for $\Delta c^{(2)}$ agree with \cite{Alishahiha:2015rta}, namely $0$ and $\frac{1}{128}$, respectively. Now, we have a formula for general $d$, not only for $\Delta c^{(2)}$, but for $\Delta c^{(3)}$ as well.

Note that $\Delta c^{(2)}$ and $\Delta c^{(3)}$ both vanish when $d=2$, that is for AdS${}_3$. In fact, we have the exact embedding for AdS${}_3$ in \eqref{eq:AdS3embed}, which yields $\Delta c = 0$ identically in this case. This happens since the AdS${}_3$ black hole, or the BTZ black hole, is a quotient of AdS${}_3$ and is thus locally the same as AdS${}_3$. Similar behavior specific to AdS${}_3$ has been observed before in the context of the complexity of formation~\cite{Chapman:2016hwi}.

\begin{figure}[t]
\centering
\includegraphics[width=0.49\textwidth]{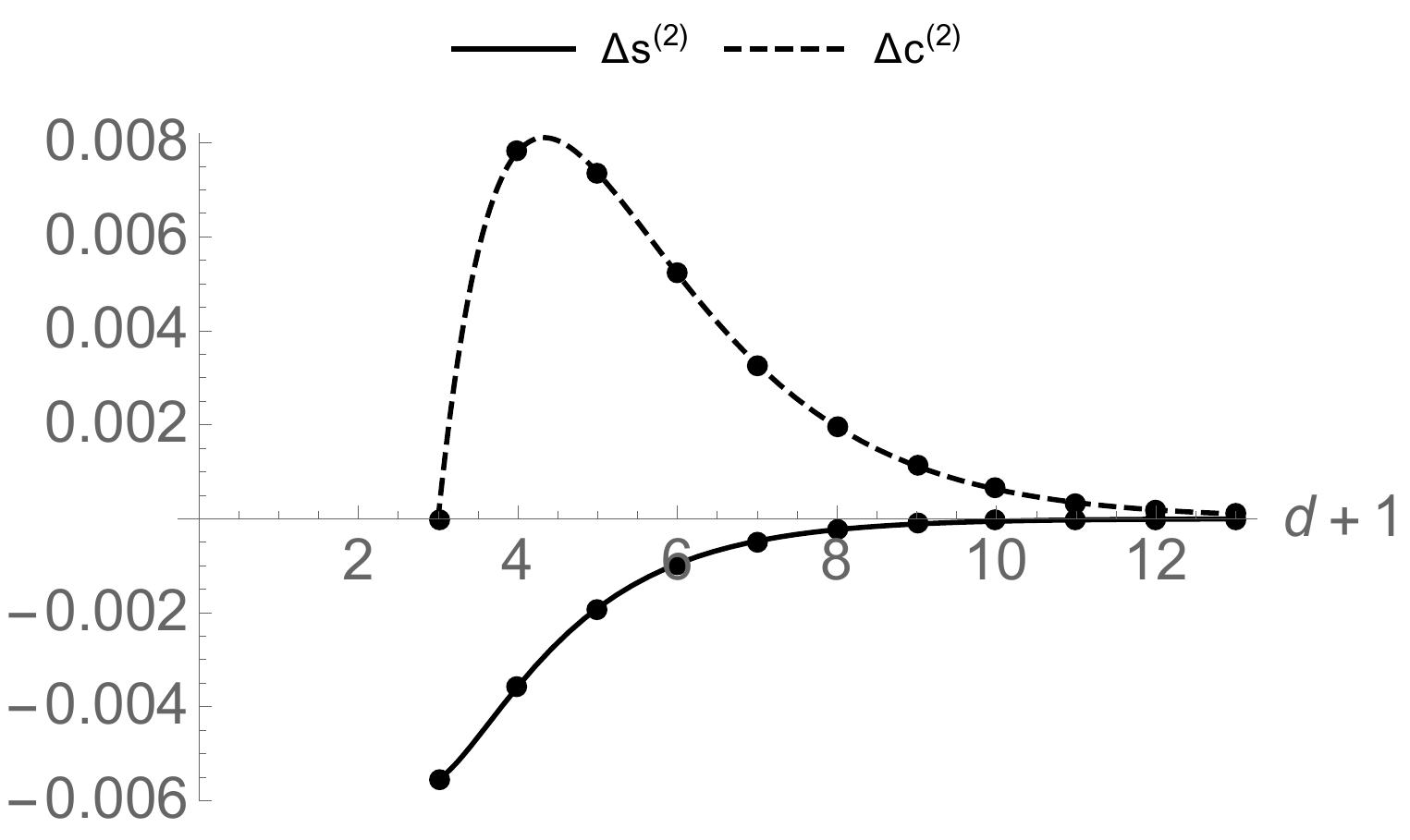}
\includegraphics[width=0.49\textwidth]{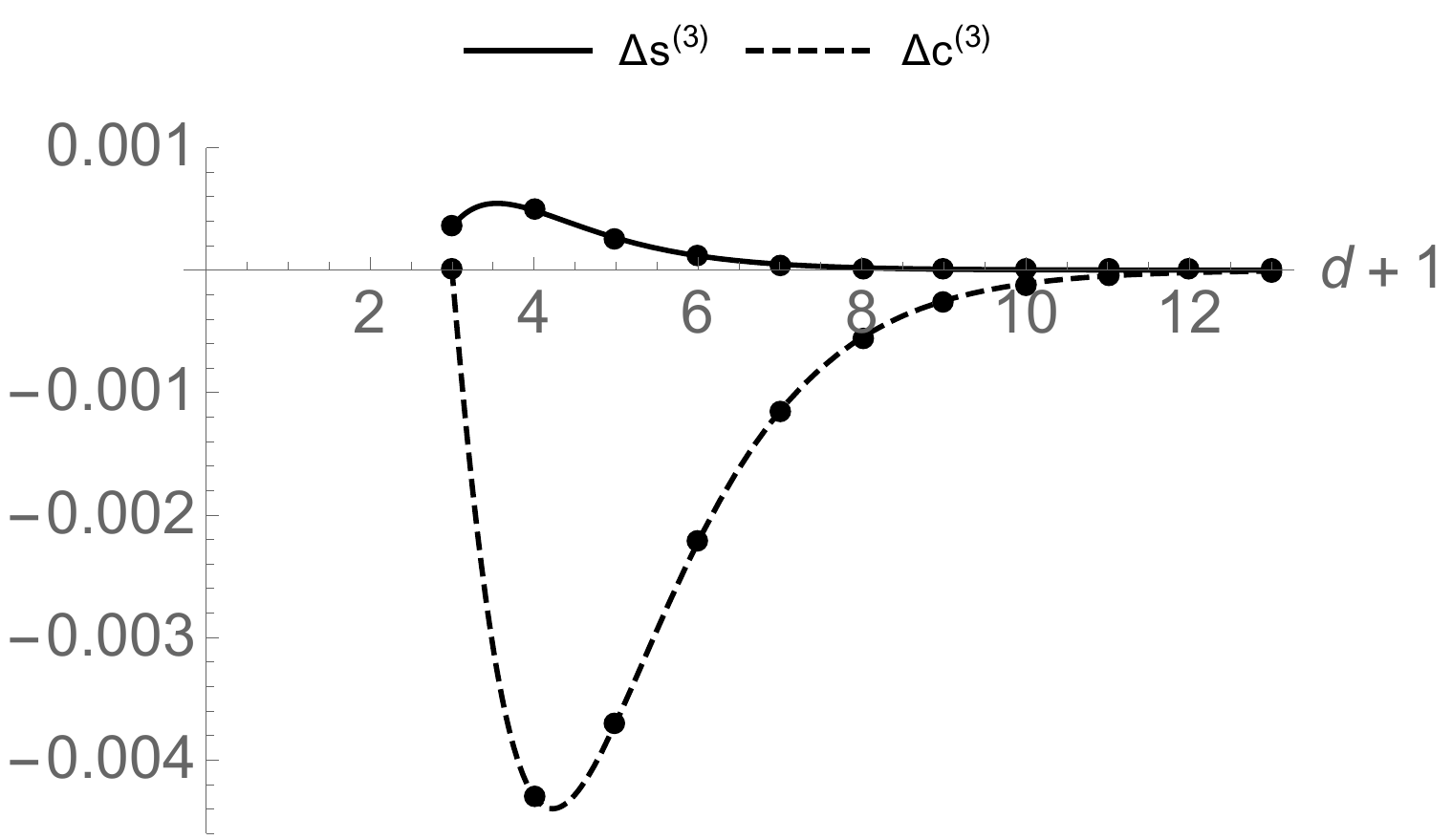}
\caption{Plots of $\Delta s$ and $\Delta c$ to second- and third-order in the uncharged black hole background. The points are explicitly calculated values. The curves are plots of the general formulae.}
\label{fig:3}
\end{figure}

We observe some interesting behavior in $\Delta s$ and $\Delta c$ up to third order in uncharged black holes. Note that $\Delta s^{(2)}$ is negative whereas $\Delta c^{(2)}$ is positive (or 0 for AdS${}_3$). At third order, the signs flip and $\Delta s^{(3)}$ is now positive whereas $\Delta c^{(3)}$ is negative (or 0 for AdS${}_3$). Only the sign of $\Delta s^{(2)}$ is constrained to be negative by the first law of entanglement \cite{Blanco:2013joa}. It is tantalizing that, $\Delta c$ appears to be of opposite sign as compared with $\Delta s$ at each order (see Figure \ref{fig:3}). We will find that this behavior continues to hold for the charged black hole. It would be interesting to see if this behavior persists in other scenarios and to higher orders and if it can be proven in general.

%%%%%%%%%%%%%%%%%%%%%%%%%%%%%%%%%%%%%%%%
% Charged Holographic Subregion Complexity
%%%%%%%%%%%%%%%%%%%%%%%%%%%%%%%%%%%%%%%%
\subsection{Charged AdS Black Hole}

In the charged AdS black hole case, we expand $\Delta c$ up to the first four orders for general $d$:
\begin{equation}
    \Delta c = \sparq^{d} \Delta c^{(1,0)} + \sparq^{2d-2} \Delta c^{(0,1)} + \sparq^{2d} \Delta c^{(2,0)} + \sparq^{3d-2} \Delta c^{(1,1)} + \cdots .
\end{equation}
Again, as in the case of entanglement entropy, we find
\begin{subequations} \label{eq:dcch}
\begin{align}
    \Delta c^{(1,0)} &=  (1+ \qzh^{2}) \Delta c^{(1)}, \label{eq:dcch1}\\
    \Delta c^{(2,0)} &=  (1+ \qzh^{2})^{2}\Delta c^{(2)}, \label{eq:dcch3} 
\end{align}
\end{subequations}
similar to \eqref{eq:yyrel} and \eqref{eq:ssrelat}.
For the newly appearing orders $(2d-2)$ and $(3d-2)$ in the charged BH case, we use embedding functions $\y_{(0,1)}( \x )$ and $\y_{(1,1)}( \x )$ derived for $d=3$, $4$, $5$ and $6$, presented in Appendix \ref{app:C}.\footnote{We are not able to generalize these embedding functions for general $d$ as the differential equations keep getting more and more complicated to solve for general $d$.} Using these embedding functions, we compute the subregion complexity changes at orders $(2d-2)$ and $(3d-2)$ for the aforementioned $d$ values. As in the uncharged case, we can separate out the dependence of these results on $\y_{(1,0)}$, which we do know for general $d$ values. Using this piece as guidance, we are able to deduce the changes of subregion complexity at order $(2d-2)$ and $(3d-2)$ for general $d$. We then checked our formula against results calculated for $d$ values higher than $6$ (up to $10$). Indeed, our formula reproduces correct results in those cases as well. The following are our expressions of $\Delta c^{(0,1)}$ and $\Delta c^{(1,1)}$:
\begin{subequations} \label{eq:dcch34}
\begin{align}
    \Delta c^{(0,1)} &=  \qzh^{2}\frac{\pi^{\frac{1}{2}}}{2^{d+1}} \frac{ (d-2) \Gamma \left(\frac{d-1}{2}\right)}{\Gamma \left(\frac{d}{2}\right)}, \label{eq:dcch2}\\
    \Delta c^{(1,1)} &= - \qzh^{2} (1+ \qzh^{2}) \frac{3 (d-1) (d-2) \Gamma \left(d-\frac{1}{2}\right) \Gamma \left(\frac{d+1}{2}\right)}{4 (d+1) \Gamma \left(\frac{3 d}{2}\right)}. \label{eq:dcch4} 
\end{align}
\end{subequations}
Looking at these results \eqref{eq:dcch}, \eqref{eq:dcch34} and comparing them with the signs of \eqref{eq:dsch}, we again see that whenever $\Delta s$ at some order is positive (negative) definite, $\Delta c$ is negative (positive) definite. Therefore, for both the uncharged black hole (pure stress tensor perturbation) and the charged black hole (mixed stress tensor and current perturbation), we observe that the change in HEE and HSC at some particular order come with opposite sign. We add that this relative minus sign between the change in HEE and HSC also holds for the leading-order result in the case of a perturbation due to a scalar of conformal dimension $\Delta$~\cite{Blanco:2013joa, Banerjee:2017qti}.

We will return to this observation in the conclusion section when we discuss the relative information that might be contained in the HEE and HSC in the case of the ball entangling region.

%%%%%%%%%%%%%%%%%%%%%%%%%%%%%%%%%%%%%%%%
% Entanglement Thermodynamics
%%%%%%%%%%%%%%%%%%%%%%%%%%%%%%%%%%%%%%%%

\section{Entanglement Thermodynamics}

The field theories dual to the uncharged and charged AdS black holes are correspondingly charged and uncharged perfect fluids with stress tensor taking the form
\begin{equation}
    T_{\mu\nu} = ( \mathcal{E} + P ) u_{\mu} u_{\nu} + P \eta_{\mu\nu},
\end{equation}
where $\mathcal{E}$ is the energy density, $P$ is the pressure and $u_{\mu}$ is the fluid velocity. In addition, the fact that the dual field theory is actually a CFT implies that
\begin{equation}
    P = \frac{\mathcal{E}}{d-1}.
\end{equation}
Matching this with the boundary metric for the case of the uncharged black hole gives the standard AdS/CFT dictionary relationship between the boundary energy density and bulk geometric data,
\begin{equation*}
    \mathcal{E} = \frac{d-1}{2} \biggl( \frac{\AdSrad}{\lp} \biggr)^{d-1} \frac{1}{\zh^d}.
\end{equation*}
Both the uncharged and charged AdS black holes correspond to perfect fluids at rest, with a fluid velocity given by $u^{\mu} = \delta_{0}^{\mu}$. The stress tensor for the charged case is $(1 + q^2 \zh^2 )$ times the stress tensor for the uncharged case in which the uncharged horizon radius is replaced with the charged one. Both cases have a constant energy density and therefore the energy contained in the ball entangling region of radius $\rad$ scales as $\rad^{d-1}$ for both cases. To be precise, for the uncharged black hole,
\begin{align} \label{eq:dE}
    \Delta E &= \int T_{00} \, d\Omega_{d-2} \, r^{d-2} dr = \frac{1}{2} \Omega_{d-2} \biggl( \frac{\AdSrad}{\lp} \biggr)^{d-1} \frac{\rad^{d-1}}{\zh^d},
\end{align}
and the charged case is the same result multiplied by $(1 + q^2 \zh^2 )$. In other words, $\Delta E$ is proportional to $\spar$ in the uncharged case and $\sparq^d$ in the charged case. This is not just a perturbative result, but is an exact one. Meanwhile, for the uncharged black hole, we expand out the entanglement entropy as
\begin{equation}
    \Delta S_E = \Delta S_{E}^{(1)} + \Delta S_{E}^{(2)} + \cdots,   
\end{equation}
where $\Delta S_{E}^{(n)}$ is a term in $\Delta S_E$ which is of order $\spar^n$.\footnote{Note that $\Delta s^{(n)}$ does not contain explicit powers of $\spar$ since that is factored out when we write \eqref{eq:dsexpand}. However, our convention here is that $\Delta S_{E}^{(n)}$ does contain an explicit factor of $\spar^n$.} We now place the explicit subscript ``$E$'' to remind the reader that we are dealing with entanglement entropy and not the usual thermodynamic entropy here. Nevertheless, the central idea of entanglement thermodynamics in \cite{Allahbakhshi:2013rda} is to make an analogy with thermodynamics and to define the \textbf{entanglement temperature} in such a way that
\begin{equation}
    \Delta E = T_E \Delta S_{E}^{(1)}.
\end{equation}
Let us make the following two observations regarding this relation:
\begin{enumerate}
    \item This is a perturbative relation that holds only at leading order;
    
    \item To extend this relation beyond leading order, one must introduce new terms because $\Delta E$ is exactly first-order while $\Delta S_E$ contains higher order corrections. Indeed, the new term would serve to cancel $T_E \Delta S_{E}^{(2)}$ at second order.
\end{enumerate}
On the other hand, as we pass to the non-perturbative regime, in which the subregion covers more and more of the entire boundary CFT, the entanglement entropy approaches the thermodynamic one, which does satisfy the laws of black hole thermodynamics. We are motivated, therefore, to try to extend the above relation at least to second order. In analogy with the usual first law, we write
\begin{equation}
    \Delta E = T_E \Delta S_{E} + W_E,
\end{equation}
where $W_E$ is some \textbf{entanglement work} analogous to thermodynamic work and encompasses the new terms mentioned in point 2 above to make the relation hold to higher order. It is important to point out that it is $\Delta E$ that appears in this relation and not $T_E \Delta \langle H \rangle$, where $H$ is the modular Hamiltonian. Firstly, the modular Hamiltonian is in general a very non-local quantity whose connection to energy is unclear. Only in the case of spherical subregions in CFTs in vacuum do we find such a direct relationship between the modular Hamiltonian and energy. Of course, that happens to be the case in study in this work, but a first law of entanglement thermodynamics ought to be more widely applicable than that. Secondly, if we were to base the first law around the modular Hamiltonian, then $W_E$ would be equivalent to $T_E S_{\text{rel}}$, where $S_{\text{rel}}$ is the relative entropy, which is always non-negative, regardless of initial and final state. Furthermore, in order for the first law to have any actual content, the work term must have an entirely distinct ontology from energy and entanglement. Otherwise, $W_E$ could simply be \textbf{defined} as $\Delta E - T_E \Delta S_E$. What could this work term be?

We do not know the answer to this question. However, we would like to point out that we are not really the first to pose the question in the first place. The question turns out to be essentially equivalent to the problem studied in \cite{Banerjee:2017qti} based off of \cite{Alishahiha:2015rta}. In fact, the authors of \cite{Banerjee:2017qti} unwittingly propose an answer to this question: the entanglement work contains a term proportional to the change in subregion complexity. To be very careful, \cite{Banerjee:2017qti} does not actually propose this directly. Instead, they propose that the Fisher information is proportional to the second-order change in the volume of the RT surface. In the context of our perturbative analysis around pure AdS, the Fisher information is just
\begin{equation} \label{eq:F}
    \mathcal{F} = \frac{d^2}{d \spar^2} \bigl( \Delta \langle H \rangle - \Delta S_E \bigr) \Bigr|_{\spar = 0} = - \frac{2}{\spar^2} \Delta S_{E}^{(2)} = \frac{\pi^{3/2}}{2^{d+2}} \frac{(d-1) \Gamma ( d+1 )}{(d+1) \Gamma \bigl( d + \frac{3}{2} \bigr)} \Omega_{d-2} \biggl( \frac{\AdSrad}{\lp} \biggr)^{d-1}.
\end{equation}
On the other hand, the change in RT volume is related to the change in subregion complexity:
\begin{equation}
    \Delta V^{(2)} = \frac{\Omega_{d-2}}{d-1} \AdSrad^d \Delta c^{(2)} = \frac{\sqrt{\pi}}{2^{d+2} (d-1)(d+1)} \frac{\Gamma \bigl( \frac{d+1}{2} \bigr)}{\Gamma \bigl( \frac{d}{2} - 1 \bigr)} \Omega_{d-1} L^d.
\end{equation}
Therefore, the proportionality constant $C_d$ defined in \cite{Banerjee:2017qti} via $\mathcal{F} = C_d \Delta V^{(2)}$ is given by\footnote{There is an ambiguity in the small parameter $\spar$ and thus an ambiguity in the definition of $\mathcal{F}$. In \cite{Lashkari:2015hha}, the derivative is taken with respect to a parameter $\mu$, which is related to our parameter $m$ by $m = 2 \mu$. Therefore, our $\mathcal{F}$ in \eqref{eq:F} is related to the Lashkari-van Raamsdonk expression by $\mathcal{F} = 4R^{2d} \mathcal{F}^{\text{LvR}}$. Note that they also set $\AdSrad = 1$ and $G_N = \frac{1}{8 \pi} \lp^{d-1} = 1$. Therefore, one finds $\mathcal{F}^{\text{LvR}} = \frac{\sqrt{\pi}}{2^{d+3}} \frac{(d-1) \Gamma (d+1)}{(d+1) \Gamma ( d+ \frac{3}{2} )} \Omega_{d-2} R^{2d}$, which is indeed $\frac{R^4}{45}$ for AdS${}_3$, as stated in \cite{Lashkari:2015hha}.}
\begin{equation} \label{eq:Cd}
    C_d = \frac{\pi (d-1)^2 \Gamma (d+1) \Gamma \bigl( \frac{d}{2} -1 \bigr)}{\spar^2 \AdSrad \lp^{d-1} \Gamma \bigl( d + \frac{3}{2} \bigr) \Gamma \bigl( \frac{d+1}{2} \bigr)}.
\end{equation}
Note that this is the first time that this coefficient has actually been computed explicitly since the expression in \cite{Banerjee:2017qti} contains a function of $d$ that was unknown until now.

Therefore, taken at face value, the suggestion in \cite{Banerjee:2017qti} is that $\Delta S_{E}^{(2)}$ is proportional to $\Delta V^{(2)}$, where $V$ is the volume of the RT surface. Therefore, though this was not its express intention, \cite{Banerjee:2017qti} suggests identifying the entanglement work with the change in the volume of the RT surface:
\begin{equation}
    \Delta E = T_E \Delta S_E + P \Delta V,
\end{equation}
where $P$ is a concomitant pressure\footnote{This pressure is unrelated to what is called \textbf{entanglement pressure} in \cite{Allahbakhshi:2013rda}.}, which is related to the coefficient $C_d$ introduced in \cite{Banerjee:2017qti} and computed in \eqref{eq:Cd} via
\begin{equation} \label{eq:pressure}
    P = T_E \frac{\spar^2}{2} C_d = \frac{d+1}{4 \pi \rad} \biggl( \frac{\rad}{\zh} \biggr)^2 C_d.
\end{equation}
The relationship between volume and complexity then says that we can equally well express the entanglement work in terms of the change in HSC,
\begin{equation} \label{eq:ESC}
    \Delta E = T_E \Delta S_E + B \Delta C,
\end{equation}
where $B = \AdSrad \lp^{d-1} P$.

In this picture, the change in HEE is morally playing the role of heat and the change in HSC is playing the role of work. In fact, the definition of complexity naturally contains within it connotations of work. It is usually defined roughly as the minimum number of unitary transformations from some prescribed collection of such transformations required to transform some particular reference state into the desired target state. It is sometimes intuitively described as the amount of ``computational power'' or ``resources'' needed to perform these operations. It is certainly not a stretch to associate this intuitive idea with some concept of work. Indeed, once a concrete and practicable definition of complexity in field theory is given, and assuming some relation like \eqref{eq:ESC} exists, then one could presumably exploit the relation to run information-theoretic periodic cycles (a.k.a. engines).

The apparent pattern that $\Delta C$ and $\Delta S$ are of opposite sign at each order is further indication that such a relation \eqref{eq:ESC} might hold. However, this cannot be the whole picture. This relation holds up to second order, but does not hold at third order. Of course we should have known that this cannot be the whole picture since it would have implied that $\Delta S$ and $\Delta C$ are not independent for the case in study. On the other hand, there is a sense in which $\Delta C$ carries more, or at least different, information than $\Delta S$, since $\Delta C$ requires more information about the embedding function than does $\Delta S$. As we have shown in Section \ref{subsec:boundary}, the $n$-th order $\Delta S$ is determined by the embedding function up to at most half that order. On the other hand, $\Delta C$ to $n$-th order depends on the embedding function up to that same order $n$. Thus, while there does appear to be a flow of information from being in the form of entanglement to subregion complexity, this transfer is not complete. From the perspective of a speculative theoretical engine, part of the work in a cycle can arise as changes in complexity and part of it can arise as something else, just as it can arise as changes in volume as well as particle number in more familiar thermodynamic cases. What other information-theoretic quantities might contribute to $W_E$ is a question worth investigating\footnote{There are a number of important works deriving Einstein's equations from entanglement, to first order (e.g.,
in \cite{Oh:2017pkr}) and then to second order
(e.g., \cite{Faulkner:2017tkh}; see also \cite{Jacobson:2015hqa}). These relate the variations of the relative entropy to bulk integrals in a formalism
developed in \cite{Hollands:2012sf}. This approach claims an exact first law of entanglement entropy from the start and it is plausible that we are rediscovering this same result perturbatively. We thank Erik Verlinde for valuable discussions in this regard. }.

%%%%%%%%%%%%%%%%%%%%%%%%%%%%%%%%%%%%%%%%
% Discussion and Outlook
%%%%%%%%%%%%%%%%%%%%%%%%%%%%%%%%%%%%%%%%

\section{Discussion and Outlook}

First, we highlight the main findings of our paper. Then, the suggestions that these results lead us to make, will follow. 
\begin{enumerate}
    \item We have computed the change in holographic entanglement entropy (HEE), $\Delta S$, and subregion complexity (HSC), $\Delta C$, for spherical entangling regions of radius $\rad$ in the background of the uncharged and charged AdS${}_{d+1}$ black holes. For the uncharged case, we have performed the calculations perturbatively in the parameter $\spar = m \rad^d$, where $m$ is the black hole mass. We find formulae as functions of $d$ for $\Delta S$ and $\Delta C$ up to third order and we also provide exact numerical results for $\Delta S^{(4)}$ in spacetime dimensions 3 to 7. For the charged case, the perturbative study has been done with respect to the small parameter $\sparq = \frac{R}{\zh}$, where $\zh$ is the charged black hole horizon radius. We compute  $\Delta S$ and $\Delta C$ up to the first four orders and have again found formulae as functions of $d$.
    
    \quad We observe that the change in entanglement entropy and subregion complexity at a particular order come with opposite signs relative to one another. This holds to all the orders we have studied for both the uncharged and charged AdS black holes. It also holds to leading order for the case of a scalar perturbation~\cite{Blanco:2013joa, Banerjee:2017qti}. This exchange in sign is mysterious from the dual field theory perspective and begs an explanation. 

    \item Another important finding of this work is the proof that the entanglement entropy change up to some order $n$ depends on the embedding function only up to the highest order less than or equal to $\frac{n}{2}$. This upper bound has not been appreciated previously, to the best of our knowledge. We hope that this allows others to push the calculations of HEE further. In addition, we note that the change in subregion complexity up to some order $n$ depends on the embedding function all the way up to that same order. We therefore gain a more quantitative sense of the information which is contained in subregion complexity but not in entanglement entropy.
\end{enumerate}

With these main results and taking inspiration from previous works, largely from \cite{Allahbakhshi:2013rda, Alishahiha:2015rta, Banerjee:2017qti}, we are lead naturally to a number of suggestions. From an information-theoretic perspective, it appears as though information is being traded between the entanglement between a boundary subregion and its complement and the complexity of the CFT state reduced to that subregion. In particular, \cite{Banerjee:2017qti} in fact inadvertently suggests that the HSC contributes a term to the first law of entanglement that is analogous to work:
\begin{equation} \label{eq:firstlaw}
    \Delta E = T_E \Delta S_E + B \Delta C,
\end{equation}
where $B$ is some known $d$-dependent quantity related to a pressure defined in \eqref{eq:pressure}. Using the closed form of the second order change in HSC, we have been able to fix the $d$-dependent constant relating this to Fisher information, as proposed previously in \cite{Banerjee:2017qti}. That a first law in the form \eqref{eq:firstlaw} does not hold in general at third order begs the existence of other information-theoretic quantities at higher orders. We hope that a more complete picture from the perspective of the information geometry will emerge from these investigations.

We note that certain modifications and generalizations to the first law of entanglement have been considered before (e.g., in \cite{Allahbakhshi:2013rda, Lokhande:2017jik}). Notably, in time-dependent scenarios (e.g., a collapsing black hole), it was found in \cite{Lokhande:2017jik} that the first law is naturally replaced by a certain linear response relation. It would be interesting to study the modifications to the linear response relations that arise at second order and if they are at all related to complexity as we have suggested here primarily for the static case.\footnote{We are grateful to Juan Pedraza for this suggestion.}

Recent works have tried to come up with various field theoretic definitions of complexity from a few different perspectives, for example, geometric and circuit complexity \cite{Jefferson:2017sdb, Chapman:2017rqy, Camargo:2018eof, Bhattacharyya:2018bbv, Ali:2018fcz, Chapman:2018hou, Ge:2019mjt} and path integral complexity \cite{Caputa:2017urj, Caputa:2017yrh}. These two perspectives have been very recently bridged in \cite{Caputa:2018kdj, Camargo:2019isp}. Using this line of study, it would be interesting to study the fidelity, primarily for free QFTs and then for holographic CFTs. It would be interesting to check whether the third and higher order expansion terms follow the relations we found in higher orders.

An intriguing line of research aims at ascribing geometry to circuit complexity and uses a \textit{cost function} to identify the optimal path in state space from a reference state to a target state~\cite{nielsen2006quantum, dowling2008geometry, nielsen2005geometric}. Recently, \cite{Bernamonti:2019zyy} has probed this idea using the complexity $=$ action conjecture. The idea here is to start with a simple target state (in this case, the CFT ground state dual to pure AdS) and perturb it slightly (in this case, by a scalar field excitation) and to glean data about derivatives of the cost function on state space from the change in complexity on the gravity side. For example, the vanishing of the first-order term in the change in complexity implies that the optimal path in state space is orthogonal to the direction in which we perturb the original target state. Information about the second derivatives of the cost function come from the second-order change in complexity, and so on. One could imagine a similar picture simply reduced to subregions. It would be interesting to see if we can use our second- and third-order results to probe the concept of a cost function on subregion-reduced state space.

Another avenue that might be interesting to explore is whether one can capture the confinement-deconfinement phase transition between the AdS black hole and AdS soliton \cite{Horowitz:1998ha} solutions by looking at changes in HEE and HSC \cite{Klebanov:2007ws, Ishihara:2012jg, Reynolds:2017jfs}. We are optimistic about this prospect since the idea that such information-theoretic quantities display interesting behavior near phase transitions has been successfully studied in a number of situations before (e.g., in \cite{Momeni:2016ekm, Zhang:2017nth, Du:2018uua}).

We hope that having the second- and third-order HSC in closed form will aid in the quest to find a purely field-theoretic definition of complexity. We hope that this will also spur further developments in the study of information-theoretic aspects of quantum gravity in general.

%%%%%%%%%%%%%%%%%%%%%%%%%%%%%%%%%%%%%%%%
% Acknowledgements
%%%%%%%%%%%%%%%%%%%%%%%%%%%%%%%%%%%%%%%%

\section*{Acknowledgements}

It is a pleasure to thank Shira Chapman, Christian Ecker, Johanna Erdmenger, Zach Fisher, Federico Galli, Avirup Ghosh, Michal Heller, Charles Melby-Thompson , Rob Myers and Erik Verlinde for helpful discussions and critical insights. A.B. would like to thank the hospitality of the TP-$3$ division of W\"urzburg University, where the idea of the work began. We wish to thank the organizers of ``Indian Strings Meeting 2018'' where part of this work was done. K.T.G. would like to thank Ziqi Yan for his invitation to the Perimeter Institute for Theoretical Physics, where some of the late stages of this research were performed. A.B. and S.R. would like to thank DAE, Govt. of India and Homi Bhabha National Institute for support. K.T.G. is grateful to W\"urzburg University and TU Dresden for a Hallwachs-R\"ontgen fellowship in connection with the Cluster of Excellence `Topology and Complexity in Quantum Matter.'

%%%%%%%%%%%%%%%%%%%%%%%%%%%%%%%%%%%%%%%%
% Appendix
%%%%%%%%%%%%%%%%%%%%%%%%%%%%%%%%%%%%%%%%

\appendix

%%%%%%%%%%%%%%%%%%%%%%%%%%%%%%%%%%%%%%%%
% Second-Order Embedding
%%%%%%%%%%%%%%%%%%%%%%%%%%%%%%%%%%%%%%%%

\section{Uncharged BH Embedding}
\label{app:A}

The second-order embedding functions $\y_2 ( \x )$ in dimensions 3 to 7 are

\begin{subequations}
\begin{align}
    \y_{2}^{\rm AdS_3} ( \x ) &= \frac{\sqrt{1 - \x^2}}{360} \bigl( 48 - 32 \x^2 + 3 \x^4 \bigr), \\[10pt]
    \y_{2}^{\rm AdS_4} ( \x ) &= \frac{\sqrt{1 - \x^2}}{4480} \bigl( 513 - 771 \x^2 + 346 \x^4 - 40 \x^6 \bigr) \notag \\
    &\quad + \frac{3}{140} \biggl( \frac{\ln \bigl( 1 + \sqrt{1 - \x^2} \bigr)}{\sqrt{1 - \x^2}} - 1 \biggr), \\[10pt]
    \y_{2}^{\rm AdS_5} ( \x ) &= \frac{( 1 - \x^2 )^{3/2}}{4200} \bigl( 376 - 592 \x^2 + 267 \x^4 - 35 \x^6 \bigr), \\[10pt]
    \y_{2}^{\rm AdS_6} ( \x ) &= \frac{\sqrt{1 - \x^2}}{66528 \x^2} \bigl( 320 - 4935 \x^2 + 18045 \x^4 - 24469 \x^6 + 15607 \x^8 - 4592 \x^{10} \notag \\
    &\hspace{2cm} + 504 \x^{12} \bigr) + \frac{10}{2079} \biggl( \frac{1}{\x^2} + 2 - \frac{3 \ln \bigl( 1 + \sqrt{1 - \x^2} \bigr)}{\sqrt{1 - \x^2}} \biggr), \\[10pt]
    \y_{2}^{\rm AdS_7} ( \x ) &= \frac{( 1 - \x^2 )^{5/2}}{168168} \bigl( 11140 -  28356 \x^2 + 25227 \x^4 - 9006 \x^6 + 1155 \x^8 \bigr).
\end{align}
\end{subequations}
The inverse relations require us to define the variable $\ys = \y / \y(0)$:
\begingroup
\allowdisplaybreaks
\begin{subequations}
\begin{align}
    \x_{2}^{\rm{AdS}{}_3} ( \ys ) &= \frac{\sqrt{1 - \ys^2}}{40} \bigl( 3 \ys^4 + 4 \ys^2 + 8 \bigr), \\[10pt]
    \x_{2}^{\rm{AdS}{}_4} ( \ys ) &= \frac{\sqrt{1 - \ys^2}}{4480 (1+\ys)^2} \bigl( 240\ys^8 + 480 \ys^7 + 639 \ys^6 + 798 \ys^5 + 634 \ys^4 + 890 \ys^3 + 1122 \ys^2 \notag \\
    &\hspace{3cm} + 1310 \ys + 703 \bigr) + \frac{3}{140 \sqrt{1 - \ys^2}} \ln \biggl( \frac{1+\ys}{2} \biggr), \\[10pt]
    \x_{2}^{\rm{AdS}{}_5} ( \ys ) &= \frac{\sqrt{1 - \ys^2}}{4200} \bigl( 175\ys^8 + 328 \ys^6 + 228 \ys^4 + 380 \ys^2 + 464 \bigr), \\[10pt]
    \x_{2}^{\rm{AdS}{}_6} ( \ys ) &= \frac{\sqrt{1 - \ys^2}}{66528 (1+\ys)^2} \bigl( 2268\ys^{12} + 4536 \ys^{11} + 6853 \ys^{10} + 9170 \ys^9 + 8046 \ys^8 + 6922 \ys^7 \notag \\
    &\quad + 5838 \ys^6 + 7526 \ys^5 + 9294 \ys^4 + 10138 \ys^3 + 11222 \ys^2 + 11666 \ys + 5353 \bigr) \notag \\
    &\quad - \frac{10}{693 \sqrt{1 - \ys^2}} \ln \biggl( \frac{1+\ys}{2} \biggr), \\[10pt]
    \x_{2}^{\rm{AdS}{}_7} ( \ys ) &= \frac{\sqrt{1 - \ys^2}}{168168} \bigl( 4851 \ys^{12} + 10332 \ys^{10} + 8196 \ys^8 + 6180 \ys^6 + 9452 \ys^4 \notag \\
    &\hspace{2.2cm} + 11168 \ys^2 + 12884 \bigr).
\end{align}
\end{subequations}
\endgroup
Note that $\y (0)$ is the turning point of the RT surface in the bulk. The point of defining $\ys$ is to impose the boundary condition $\x ( \ys = 1 ) = 0$.

We will present the third-order embedding functions only in the $\y ( \x )$ parametrization:
\begin{subequations}
\begin{align}
    \y_{3}^{\rm AdS_3} ( \x ) &= \frac{\sqrt{1- \x^2}}{15120} ( 3 \x^6 - 46 \x^4 + 584 \x^2 - 816 ), \\
    \y_{3}^{\rm AdS_4} ( \x ) &= \frac{3}{280} ( 1 - \x^2 )^{3/2} + \frac{(1 - \x^2 )}{2508800} ( 1400 \x^{8} - 13055 \x^6 + 89470 \x^4 - 204 924 \x^2 + 128544 ) \notag \\
    &\quad + \frac{3}{560} ( 3 \x^2 -4 ) \ln \bigl( 1+ \sqrt{1 - \x^2} \bigr) - \frac{3}{280} \biggl( \frac{\ln \bigl( 1+ \sqrt{1 - \x^2} \bigr)}{\sqrt{1 - \x^2}} - 1 \biggr), \\
    \y_{3}^{\rm AdS_5} ( \x ) &= \frac{( 1 - \x^2 )^{3/2}}{30030000} \bigl( 21175 \x^{10} - 193940 \x^8 + 1106251 \x^6 - 2993238 \x^4 \notag \\
    &\quad + 3441368 \x^2 - 1405296 \bigr), \\
    \y_{3}^{\rm AdS_6} ( \x ) &= \frac{1 - \x^2}{684972288 \x^2} ( 513513 \x^{16} - 5481333 \x^{14} + 32079432 \x^{12} - 109571268 \x^{10} \notag \\
    &\quad + 211009892 \x^8 - 225264756 \x^6 + 128106720 \x^4 - 37981640 \x^2 + 2196480 ) \notag \\
    &\quad - \frac{10 \sqrt{1 - \x^2}}{6237} ( 3 \x^6 - 4 \x^4 - 7 \x^2 + 2 ) + \frac{10 ( 1 - \x^2 )}{2079} ( 3- 2 \x^2 ) \ln \bigl( 1+ \sqrt{1 - \x^2} \bigr) \notag \\
    &\quad + \frac{40}{2079} \biggl( \frac{\ln \bigl( 1+ \sqrt{1 - \x^2} \bigr)}{\sqrt{1 - \x^2}} - 1 \biggr), \\
    \y_{3}^{\rm AdS_7} ( \x ) &= \frac{( 1 - \x^2 )^{5/2}}{760455696} \bigl( 569415 \x^{14} - 5926650 \x^{12} + 32824206 \x^{10} - 109100880 \x^8 \notag \\
    &\quad + 208361675 \x^6 - 222638554 \x^4 + 124255928 \x^2 - 28455732 \bigr).
\end{align}
\end{subequations}
We are able to give the exact embedding for the AdS${}_3$ BH. This is given by
\begin{equation} \label{eq:AdS3embed}
    \y^{\rm AdS_3} ( \x ) = \frac{1}{\sqrt{\spar}} \sqrt{1 - \frac{\cosh^2 \bigl( \x \sqrt{\spar} \bigr)}{\cosh^2 \bigl( \sqrt{\spar} \bigr)}}.
\end{equation}
The perturbative expansion of this around $\spar = 0$ up to third order precisely gives the AdS${}_3$ second- and third-order results given above.
 
%%%%%%%%%%%%%%%%%%%%%%%%%%%%%%%%%%%%%%%%
% Charged BH Embeddings
%%%%%%%%%%%%%%%%%%%%%%%%%%%%%%%%%%%%%%%% 
 \section{Charged BH Embedding}
\label{app:C}
The $(2d-2)$-order embedding functions $\y_{(0,1)} ( \x )$ in dimensions 4 to 7 are
\begin{subequations}
\begin{align}
    \y_{(0,1)}^{\rm AdS_4} ( \x ) &= \frac{1}{30} \qzh^2 \biggl[ \sqrt{1 - \x^2} ( 3 \x^4 - 8 \x^2 + 9 ) + 8 \biggl( \frac{\ln \bigl( 1 + \sqrt{1 - \x^2} \bigr)}{\sqrt{1 - \x^2}} - 1 \biggr) \biggr], \\[10pt]
    \y_{(0,1)}^{\rm AdS_5} ( \x ) &= \frac{1}{70} \qzh^{2} \left( 1 - \x^{2} \right)^{\frac{3}{2}} \left( 5 \x^{4} - 13 \x^{2} + 11\right) \\[10pt]
    \y_{(0,1)}^{\rm AdS_6} ( \x ) &= \frac{1}{315} \qzh^2 \biggl[ \frac{\sqrt{1 - \x^2}}{2 \x^2} \bigl( 35 \x^{10} - 160 \x^8 + 286 \x^6 - 240 \x^4 + 63 \x^2 -32 \bigr) \notag \\
    &\quad + 16 \biggl( \frac{1}{\x^2} + 2 - \frac{3 \ln \bigl( 1 + \sqrt{1 - \x^2} \bigr)}{\sqrt{1 - \x^2}} \biggr) \biggr], \\[10pt]
    \y_{(0,1)}^{\rm AdS_7} ( \x ) &= -\frac{\qzh^{2}}{4158}(1 - \x^{2})^{\frac{5}{2}} \bigl( 189 \x^{6} - 672 \x^{4} + 852 \x^{2} - 409 \bigr). 
\end{align}
\end{subequations}
The $(3d-2)$-order embedding functions $\y_{(1,1)} ( \x )$ in dimensions 4 to 7 are
\begin{subequations}
\begin{align}
    \y_{(1,1)}^{\rm AdS_4} ( \x ) &= \frac{\qzh^2 ( 1 + \qzh^2 )}{15} \biggl[ 2 ( 1 - \x^2 )^{\frac{3}{2}} + \frac{1- \x^2}{480} \bigl( 135 \x^6 - 990 \x^4 + 2328 \x^2 - 1568 \bigr) \notag \\
    &\quad + 2 \biggl( 1 - \frac{\ln \bigl( 1 + \sqrt{1 - \x^2} \bigr)}{\sqrt{1 - \x^2}} \biggr) + (3 \x^2 - 4 ) \ln \bigl( 1 + \sqrt{1 - \x^2} \bigr) \biggr], \\[10pt]
    \y_{(1,1)}^{\rm AdS_5} ( \x ) &= -\frac{\qzh^{2} ( 1 + \qzh^{2} )}{23100} \left(1 - \x^{2} \right)^{\frac{3}{2}} \bigl( 405 \x^{8} - 3020 \x^{6} + 7833 \x^{4} - 9199 \x^{2} + 4261 \bigr) \\[10pt]
    \y_{(1,1)}^{\rm AdS_6} ( \x ) &= \frac{\qzh^2 (1 + \qzh^2 )}{945} \biggl[ - \frac{16 \sqrt{ 1 - \x^2}}{\x^2} \bigl( 3 \x^6 - 4 \x^4 - 7 \x^2 + 2 \bigr)  \notag \\
    &\quad - \frac{1- \x^2}{9240} \bigl( 138600 \x^{12} - 1258950 \x^{10} + 4551330 \x^8 - 8660015 \x^6 + 9241635 \x^4 \notag \\
    &\qquad\qquad - 5719176 \x^2 + 2593616 \bigr) + 48 ( 1 - \x^2 ) ( 3 - 2 \x^2 ) \ln \bigl( 1 + \sqrt{ 1 - \x^2 } \bigr) \notag \\
    &\quad + 32 \biggl( \frac{1}{\x^2} - 7 + \frac{6 \ln \bigl( 1 + \sqrt{1 - \x^2} \bigr)}{\sqrt{ 1- \x^2}} \biggr) \biggr], \\[10pt]
    \y_{(1,1)}^{\rm AdS_7} ( \x ) &= -\frac{\qzh^{2} \left( 1 + \qzh^2 \right)}{989604} \left(1 - \x^{2} \right)^{\frac{5}{2}} \Bigl( 14175 \x^{12} - 125832 \x^{10}
    \notag \\
    &\hspace{1cm} + 447405 \x^{8} - 837755 \x^{6} + 887386 \x^{4} - 510857 \x^{2} + 127526 \Bigr). 
\end{align}
\end{subequations}

As we can see by looking at these embedding functions, $\y_{(0,1)}( \x )$ always carries the pre-factor $p^{2}$ with it whereas $\y_{(1,1)}( \x )$, being a mixture of orders $d$ and $(2d-2)$, consistently carries a pre-factor $\qzh^{2}( 1 + \qzh^{2} )$ along with it. In our plots of the embedding functions, we plot the functions of x apart from these pre-factors. If one wishes to get the exact rescaled plots for some particular $\qzh$, these plots will be rescaled with these respective pre-factors.

We also have the inverse forms of these embedding functions. But we do not present them as they are very big expressions and we do not necessarily need them. The results were reproduced using the inverse embedding functions as well and were unchanged.
 
%%%%%%%%%%%%%%%%%%%%%%%%%%%%%%%%%%%%%%%%
% Fourth-Order Change in Entanglement Entropy
%%%%%%%%%%%%%%%%%%%%%%%%%%%%%%%%%%%%%%%%

\section{Fourth-Order Change in Entanglement Entropy}
\label{app:ct}

The fourth-order change in HEE is given by
\begin{align}
    \Delta s^{(4)} &= ( s_{0,1111} + s_{1,111} + s_{2,11} + s_{3,1} + s_{4,0} ) + ( s_{0,22} + s_{0,112} + s_{1,12} + s_{2,2} ) \notag \\
    &\quad + ( s_{0,13} + s_{1,3} ) + s_{0,4} \notag \\
    &= ( s_{0,1111} + s_{1,111} + s_{2,11} + s_{3,1} + s_{4,0} ) - s_{0,22},
\end{align}
where the contribution of $\y_4$ and $\y_3$ vanish by virtue of the Euler-Lagrange equations for $\y_0$ and $\y_1$, respectively. Furthermore, and the contribution of $\y_2$ simplifies significantly by virtue of the Euler-Lagrange equation for $\y_2$ itself. These simplifications are discussed and proven in Section \ref{subsec:boundary}.

Despite the simplifications, this still depends explicitly on $\y_2$. Since we are unable to perform the requisite integrals using the general form of $\y_2$ as a function of $d$ given in \eqref{eq:y2}, we have to infer the general formula for $\Delta s^{(4)}$ from results at specific values of $d$. In general, this is a difficult task and we cannot yet give a general formula for $\Delta s^{(4)}$. Nevertheless, we give the values of $\Delta s^{(4)}$ for AdS${}_3$ to AdS${}_7$ below.
\begin{center}
\begin{tabular}{ l  c}
\hline
AdS${}_{d+1}$ & $\Delta S^{(4)}$ $\bigl($in units of $2 \pi \Omega_{d-2} \bigl( \frac{\AdSrad}{\lp} \bigr)^{d-1} m^4 \rad^{4d} \bigr)$ \\
\hline
  AdS${}_3$ & $- \frac{1}{37800}$ \\
  AdS${}_4$ & $\frac{643689}{3139136000} - \frac{9 \ln 2}{19600}$ \\
  AdS${}_5$ & $- \frac{213784}{3350221875}$ \\
  AdS${}_6$ & $\frac{5(-824827123 + 931170240 \ln 2)}{33539518244232}$ \\
  AdS${}_7$ & $- \frac{54651392}{5471241090315}$ \\
\hline
\end{tabular}
\end{center}
Note that, just like $\Delta S^{(2)}$ and $\Delta S^{(3)}$, we find that $\Delta S^{(4)}$ is also of fixed sign. In this case, $\Delta S^{(4)}$ is negative. This suggests that the change in HEE is of fixed sign at each order and it appears to alternate from positive to negative at odd and even orders, respectively. There may be interesting physics underlying this observation, which we postpone to future investigation.

%%%%%%%%%%%%%%%%%%%%%%%%%%%%%%%%%%%%%%%%
% References
%%%%%%%%%%%%%%%%%%%%%%%%%%%%%%%%%%%%%%%%

%\providecommand{\href}[2]{#2}\begingroup\raggedright\begin{thebibliography}{99}
\bibliographystyle{JHEP}
\bibliography{HEEHSCbib.bib}% Produces the bibliography via BibTeX.

\providecommand{\href}[2]{#2}\begingroup\raggedright\begin{thebibliography}{10}

\bibitem{Maldacena:1997re}
J.~M. Maldacena, \emph{The large n limit of superconformal field theories and
  supergravity}, \href{http://dx.doi.org/10.1023/A:1026654312961,
  10.4310/ATMP.1998.v2.n2.a1}{\emph{Int. J. Theor. Phys.} {\bf 38} (1999)
  1113--1133}, [\href{https://arxiv.org/abs/hep-th/9711200}{{\tt
  hep-th/9711200}}].

\bibitem{Aharony:1999ti}
O.~Aharony, S.~S. Gubser, J.~M. Maldacena, H.~Ooguri and Y.~Oz, \emph{Large n
  field theories, string theory and gravity},
  \href{http://dx.doi.org/10.1016/S0370-1573(99)00083-6}{\emph{Phys. Rept.}
  {\bf 323} (2000) 183--386}, [\href{https://arxiv.org/abs/hep-th/9905111}{{\tt
  hep-th/9905111}}].

\bibitem{Kovtun:2004de}
P.~Kovtun, D.~T. Son and A.~O. Starinets, \emph{Viscosity in strongly
  interacting quantum field theories from black hole physics},
  \href{http://dx.doi.org/10.1103/PhysRevLett.94.111601}{\emph{Phys. Rev.
  Lett.} {\bf 94} (2005) 111601},
  [\href{https://arxiv.org/abs/hep-th/0405231}{{\tt hep-th/0405231}}].

\bibitem{Luzum:2008cw}
M.~Luzum and P.~Romatschke, \emph{Conformal relativistic viscous hydrodynamics:
  Applications to rhic results at $\sqrt{s_{NN}} = 200$ gev},
  \href{http://dx.doi.org/10.1103/PhysRevC.78.034915,
  10.1103/PhysRevC.79.039903}{\emph{Phys. Rev.} {\bf C78} (2008) 034915},
  [\href{https://arxiv.org/abs/0804.4015}{{\tt 0804.4015}}].

\bibitem{Rangamani:2009xk}
M.~Rangamani, \emph{Gravity and hydrodynamics: Lectures on the fluid-gravity
  correspondence},
  \href{http://dx.doi.org/10.1088/0264-9381/26/22/224003}{\emph{Class. Quant.
  Grav.} {\bf 26} (2009) 224003}, [\href{https://arxiv.org/abs/0905.4352}{{\tt
  0905.4352}}].

\bibitem{Barbon:2009zza}
J.~L.~F. Barbon, \emph{{Black holes, information and holography}},
  \href{http://dx.doi.org/10.1063/1.3141233}{\emph{AIP Conf. Proc.} {\bf 1122}
  (2009) 12--18}.

\bibitem{Almheiri:2012rt}
A.~Almheiri, D.~Marolf, J.~Polchinski and J.~Sully, \emph{Black holes:
  Complementarity or firewalls?},
  \href{http://dx.doi.org/10.1007/JHEP02(2013)062}{\emph{JHEP} {\bf 02} (2013)
  062}, [\href{https://arxiv.org/abs/1207.3123}{{\tt 1207.3123}}].

\bibitem{nielsenbook}
M.~Nielsen and I.~Chuang, \emph{Quantum computation and quantum information}.
\newblock Cambridge University Press, 2010.

\bibitem{Ryu:2006bv}
S.~Ryu and T.~Takayanagi, \emph{Holographic derivation of entanglement entropy
  from ads/cft},
  \href{http://dx.doi.org/10.1103/PhysRevLett.96.181602}{\emph{Phys. Rev.
  Lett.} {\bf 96} (2006) 181602},
  [\href{https://arxiv.org/abs/hep-th/0603001}{{\tt hep-th/0603001}}].

\bibitem{Ryu:2006ef}
S.~Ryu and T.~Takayanagi, \emph{{Aspects of Holographic Entanglement Entropy}},
  \href{http://dx.doi.org/10.1088/1126-6708/2006/08/045}{\emph{JHEP} {\bf 08}
  (2006) 045}, [\href{https://arxiv.org/abs/hep-th/0605073}{{\tt
  hep-th/0605073}}].

\bibitem{Bombelli:1986rw}
L.~Bombelli, R.~K. Koul, J.~Lee and R.~D. Sorkin, \emph{A quantum source of
  entropy for black holes},
  \href{http://dx.doi.org/10.1103/PhysRevD.34.373}{\emph{Phys. Rev.} {\bf D34}
  (1986) 373--383}.

\bibitem{Srednicki:1993im}
M.~Srednicki, \emph{Entropy and area},
  \href{http://dx.doi.org/10.1103/PhysRevLett.71.666}{\emph{Phys. Rev. Lett.}
  {\bf 71} (1993) 666--669}, [\href{https://arxiv.org/abs/hep-th/9303048}{{\tt
  hep-th/9303048}}].

\bibitem{Holzhey:1994we}
C.~Holzhey, F.~Larsen and F.~Wilczek, \emph{Geometric and renormalized entropy
  in conformal field theory},
  \href{http://dx.doi.org/10.1016/0550-3213(94)90402-2}{\emph{Nucl. Phys.} {\bf
  B424} (1994) 443--467}, [\href{https://arxiv.org/abs/hep-th/9403108}{{\tt
  hep-th/9403108}}].

\bibitem{Calabrese:2004eu}
P.~Calabrese and J.~L. Cardy, \emph{Entanglement entropy and quantum field
  theory}, \href{http://dx.doi.org/10.1088/1742-5468/2004/06/P06002}{\emph{J.
  Stat. Mech.} {\bf 0406} (2004) P06002},
  [\href{https://arxiv.org/abs/hep-th/0405152}{{\tt hep-th/0405152}}].

\bibitem{Eisert:2008ur}
J.~Eisert, M.~Cramer and M.~B. Plenio, \emph{Area laws for the entanglement
  entropy - a review},
  \href{http://dx.doi.org/10.1103/RevModPhys.82.277}{\emph{Rev. Mod. Phys.}
  {\bf 82} (2010) 277--306}, [\href{https://arxiv.org/abs/0808.3773}{{\tt
  0808.3773}}].

\bibitem{Nishioka:2009un}
T.~Nishioka, S.~Ryu and T.~Takayanagi, \emph{Holographic entanglement entropy:
  An overview},
  \href{http://dx.doi.org/10.1088/1751-8113/42/50/504008}{\emph{J. Phys.} {\bf
  A42} (2009) 504008}, [\href{https://arxiv.org/abs/0905.0932}{{\tt
  0905.0932}}].

\bibitem{Takayanagi:2012kg}
T.~Takayanagi, \emph{Entanglement entropy from a holographic viewpoint},
  \href{http://dx.doi.org/10.1088/0264-9381/29/15/153001}{\emph{Class. Quant.
  Grav.} {\bf 29} (2012) 153001}, [\href{https://arxiv.org/abs/1204.2450}{{\tt
  1204.2450}}].

\bibitem{Witten:2018lha}
E.~Witten, \emph{Aps medal for exceptional achievement in research: Invited
  article on entanglement properties of quantum field theory},
  \href{http://dx.doi.org/10.1103/RevModPhys.90.045003}{\emph{Rev. Mod. Phys.}
  {\bf 90} (2018) 045003}, [\href{https://arxiv.org/abs/1803.04993}{{\tt
  1803.04993}}].

\bibitem{Lewkowycz:2013nqa}
A.~Lewkowycz and J.~Maldacena, \emph{Generalized gravitational entropy},
  \href{http://dx.doi.org/10.1007/JHEP08(2013)090}{\emph{JHEP} {\bf 08} (2013)
  090}, [\href{https://arxiv.org/abs/1304.4926}{{\tt 1304.4926}}].

\bibitem{Faulkner:2013ana}
T.~Faulkner, A.~Lewkowycz and J.~Maldacena, \emph{Quantum corrections to
  holographic entanglement entropy},
  \href{http://dx.doi.org/10.1007/JHEP11(2013)074}{\emph{JHEP} {\bf 11} (2013)
  074}, [\href{https://arxiv.org/abs/1307.2892}{{\tt 1307.2892}}].

\bibitem{Engelhardt:2014gca}
N.~Engelhardt and A.~C. Wall, \emph{{Quantum Extremal Surfaces: Holographic
  Entanglement Entropy beyond the Classical Regime}},
  \href{http://dx.doi.org/10.1007/JHEP01(2015)073}{\emph{JHEP} {\bf 01} (2015)
  073}, [\href{https://arxiv.org/abs/1408.3203}{{\tt 1408.3203}}].

\bibitem{Jafferis:2015del}
D.~L. Jafferis, A.~Lewkowycz, J.~Maldacena and S.~J. Suh, \emph{Relative
  entropy equals bulk relative entropy},
  \href{http://dx.doi.org/10.1007/JHEP06(2016)004}{\emph{JHEP} {\bf 06} (2016)
  004}, [\href{https://arxiv.org/abs/1512.06431}{{\tt 1512.06431}}].

\bibitem{Dong:2017xht}
X.~Dong and A.~Lewkowycz, \emph{{Entropy, Extremality, Euclidean Variations,
  and the Equations of Motion}},
  \href{http://dx.doi.org/10.1007/JHEP01(2018)081}{\emph{JHEP} {\bf 01} (2018)
  081}, [\href{https://arxiv.org/abs/1705.08453}{{\tt 1705.08453}}].

\bibitem{Belin:2018juv}
A.~Belin, N.~Iqbal and S.~F. Lokhande, \emph{Bulk entanglement entropy in
  perturbative excited states},
  \href{http://dx.doi.org/10.21468/SciPostPhys.5.3.024}{\emph{SciPost Phys.}
  {\bf 5} (2018) 024}, [\href{https://arxiv.org/abs/1805.08782}{{\tt
  1805.08782}}].

\bibitem{Ghosh:2017ygi}
A.~Ghosh and R.~Mishra, \emph{Inhomogeneous jacobi equation for minimal
  surfaces and perturbative change in holographic entanglement entropy},
  \href{http://dx.doi.org/10.1103/PhysRevD.97.086012}{\emph{Phys. Rev.} {\bf
  D97} (2018) 086012}, [\href{https://arxiv.org/abs/1710.02088}{{\tt
  1710.02088}}].

\bibitem{He:2014lfa}
S.~He, J.-R. Sun and H.-Q. Zhang, \emph{On holographic entanglement entropy
  with second order excitations},
  \href{http://dx.doi.org/10.1016/j.nuclphysb.2018.01.015}{\emph{Nucl. Phys.}
  {\bf B928} (2018) 160--181}, [\href{https://arxiv.org/abs/1411.6213}{{\tt
  1411.6213}}].

\bibitem{Blanco:2013joa}
D.~D. Blanco, H.~Casini, L.-Y. Hung and R.~C. Myers, \emph{Relative entropy and
  holography}, \href{http://dx.doi.org/10.1007/JHEP08(2013)060}{\emph{JHEP}
  {\bf 08} (2013) 060}, [\href{https://arxiv.org/abs/1305.3182}{{\tt
  1305.3182}}].

\bibitem{Bhattacharya:2012mi}
J.~Bhattacharya, M.~Nozaki, T.~Takayanagi and T.~Ugajin, \emph{Thermodynamical
  property of entanglement entropy for excited states},
  \href{http://dx.doi.org/10.1103/PhysRevLett.110.091602}{\emph{Phys. Rev.
  Lett.} {\bf 110} (2013) 091602}, [\href{https://arxiv.org/abs/1212.1164}{{\tt
  1212.1164}}].

\bibitem{Susskind:2014rva}
L.~Susskind, \emph{{Computational Complexity and Black Hole Horizons}},
  \href{http://dx.doi.org/10.1002/prop.201500093,
  10.1002/prop.201500092}{\emph{Fortsch. Phys.} {\bf 64} (2016) 44--48},
  [\href{https://arxiv.org/abs/1403.5695}{{\tt 1403.5695}}].

\bibitem{Brown:2015bva}
A.~R. Brown, D.~A. Roberts, L.~Susskind, B.~Swingle and Y.~Zhao,
  \emph{Holographic complexity equals bulk action?},
  \href{http://dx.doi.org/10.1103/PhysRevLett.116.191301}{\emph{Phys. Rev.
  Lett.} {\bf 116} (2016) 191301},
  [\href{https://arxiv.org/abs/1509.07876}{{\tt 1509.07876}}].

\bibitem{Brown:2015lvg}
A.~R. Brown, D.~A. Roberts, L.~Susskind, B.~Swingle and Y.~Zhao,
  \emph{Complexity, action, and black holes},
  \href{http://dx.doi.org/10.1103/PhysRevD.93.086006}{\emph{Phys. Rev.} {\bf
  D93} (2016) 086006}, [\href{https://arxiv.org/abs/1512.04993}{{\tt
  1512.04993}}].

\bibitem{Alishahiha:2015rta}
M.~Alishahiha, \emph{Holographic complexity},
  \href{http://dx.doi.org/10.1103/PhysRevD.92.126009}{\emph{Phys. Rev.} {\bf
  D92} (2015) 126009}, [\href{https://arxiv.org/abs/1509.06614}{{\tt
  1509.06614}}].

\bibitem{Uhlmann:1975kt}
A.~Uhlmann, \emph{The transition probability for states of *-algebras.},
  {\emph{Annalen Phys.} {\bf 42} (1985) 524}.

\bibitem{Hayashi}
M.~Hayashi, \emph{Quantum Information: an Introduction}.
\newblock Springer-Verlag, 2006.

\bibitem{Petz}
D.~Petz and C.~Ghinea in \emph{Quantum probability and related topics},
  no.~261.

\bibitem{Lashkari:2015hha}
N.~Lashkari and M.~Van~Raamsdonk, \emph{Canonical energy is quantum fisher
  information}, \href{http://dx.doi.org/10.1007/JHEP04(2016)153}{\emph{JHEP}
  {\bf 04} (2016) 153}, [\href{https://arxiv.org/abs/1508.00897}{{\tt
  1508.00897}}].

\bibitem{MIyaji:2015mia}
M.~Miyaji, T.~Numasawa, N.~Shiba, T.~Takayanagi and K.~Watanabe, \emph{Distance
  between quantum states and gauge-gravity duality},
  \href{http://dx.doi.org/10.1103/PhysRevLett.115.261602}{\emph{Phys. Rev.
  Lett.} {\bf 115} (2015) 261602},
  [\href{https://arxiv.org/abs/1507.07555}{{\tt 1507.07555}}].

\bibitem{Alishahiha:2017cuk}
M.~Alishahiha and A.~Faraji~Astaneh, \emph{{Holographic Fidelity
  Susceptibility}},
  \href{http://dx.doi.org/10.1103/PhysRevD.96.086004}{\emph{Phys. Rev.} {\bf
  D96} (2017) 086004}, [\href{https://arxiv.org/abs/1705.01834}{{\tt
  1705.01834}}].

\bibitem{Banerjee:2017qti}
S.~Banerjee, J.~Erdmenger and D.~Sarkar, \emph{Connecting fisher information to
  bulk entanglement in holography},
  \href{http://dx.doi.org/10.1007/JHEP08(2018)001}{\emph{JHEP} {\bf 08} (2018)
  001}, [\href{https://arxiv.org/abs/1701.02319}{{\tt 1701.02319}}].

\bibitem{Abt:2017pmf}
R.~Abt, J.~Erdmenger, H.~Hinrichsen, C.~M. Melby-Thompson, R.~Meyer, C.~Northe
  et~al., \emph{Topological complexity in ads$_3$/cft$_2$},
  \href{http://dx.doi.org/10.1002/prop.201800034}{\emph{Fortsch. Phys.} {\bf
  66} (2018) 1800034}, [\href{https://arxiv.org/abs/1710.01327}{{\tt
  1710.01327}}].

\bibitem{Abt:2018ywl}
R.~Abt, J.~Erdmenger, M.~Gerbershagen, C.~M. Melby-Thompson and C.~Northe,
  \emph{Holographic subregion complexity from kinematic space},
  \href{http://dx.doi.org/10.1007/JHEP01(2019)012}{\emph{JHEP} {\bf 01} (2019)
  012}, [\href{https://arxiv.org/abs/1805.10298}{{\tt 1805.10298}}].

\bibitem{Bernamonti:2019zyy}
A.~Bernamonti, F.~Galli, J.~Hernandez, R.~C. Myers, S.-M. Ruan and J.~Sim\'on,
  \emph{{The First Law of Complexity}},
  \href{https://arxiv.org/abs/1903.04511}{{\tt 1903.04511}}.

\bibitem{nielsen2005geometric}
M.~A. Nielsen, \emph{A geometric approach to quantum circuit lower bounds},
  {\emph{arXiv preprint quant-ph/0502070} (2005) }.

\bibitem{nielsen2006quantum}
M.~A. Nielsen, M.~R. Dowling, M.~Gu and A.~C. Doherty, \emph{Quantum
  computation as geometry}, {\emph{Science} {\bf 311} (2006) 1133--1135}.

\bibitem{OBannon:2016exv}
A.~O'Bannon, J.~Probst, R.~Rodgers and C.~F. Uhlemann, \emph{First law of
  entanglement rates from holography},
  \href{http://dx.doi.org/10.1103/PhysRevD.96.066028}{\emph{Phys. Rev.} {\bf
  D96} (2017) 066028}, [\href{https://arxiv.org/abs/1612.07769}{{\tt
  1612.07769}}].

\bibitem{Jaksland:2017nqx}
R.~Jaksland, \emph{A review of the holographic relation between linearized
  gravity and the first law of entanglement entropy},
  \href{https://arxiv.org/abs/1711.10854}{{\tt 1711.10854}}.

\bibitem{Allahbakhshi:2013rda}
D.~Allahbakhshi, M.~Alishahiha and A.~Naseh, \emph{Entanglement
  thermodynamics}, \href{http://dx.doi.org/10.1007/JHEP08(2013)102}{\emph{JHEP}
  {\bf 08} (2013) 102}, [\href{https://arxiv.org/abs/1305.2728}{{\tt
  1305.2728}}].

\bibitem{Taylor:2016aoi}
M.~Taylor and W.~Woodhead, \emph{Renormalized entanglement entropy},
  \href{http://dx.doi.org/10.1007/JHEP08(2016)165}{\emph{JHEP} {\bf 08} (2016)
  165}, [\href{https://arxiv.org/abs/1604.06808}{{\tt 1604.06808}}].

\bibitem{siklos}
S.~Siklos, \emph{The papperitz equation},  {MAE207 Applications of Complex
  Analysis Course Notes}, Department of Mathematics and Theoretical Physics,
  University of Cambridge,
  \url{http://www.damtp.cam.ac.uk/user/stcs/courses/fcm/handouts/papperitz.pdf},
  2007.

\bibitem{Albash:2010mv}
T.~Albash and C.~V. Johnson, \emph{{Evolution of Holographic Entanglement
  Entropy after Thermal and Electromagnetic Quenches}},
  \href{http://dx.doi.org/10.1088/1367-2630/13/4/045017}{\emph{New J. Phys.}
  {\bf 13} (2011) 045017}, [\href{https://arxiv.org/abs/1008.3027}{{\tt
  1008.3027}}].

\bibitem{Ecker:2018jgh}
C.~Ecker, \emph{{Entanglement Entropy from Numerical Holography}}.
\newblock PhD thesis, Vienna, Tech. U., 2018-09.
\newblock \href{https://arxiv.org/abs/1809.05529}{{\tt 1809.05529}}.

\bibitem{Ben-Ami:2016qex}
O.~Ben-Ami and D.~Carmi, \emph{On volumes of subregions in holography and
  complexity}, \href{http://dx.doi.org/10.1007/JHEP11(2016)129}{\emph{JHEP}
  {\bf 11} (2016) 129}, [\href{https://arxiv.org/abs/1609.02514}{{\tt
  1609.02514}}].

\bibitem{Chapman:2016hwi}
S.~Chapman, H.~Marrochio and R.~C. Myers, \emph{{Complexity of Formation in
  Holography}}, \href{http://dx.doi.org/10.1007/JHEP01(2017)062}{\emph{JHEP}
  {\bf 01} (2017) 062}, [\href{https://arxiv.org/abs/1610.08063}{{\tt
  1610.08063}}].

\bibitem{Oh:2017pkr}
E.~Oh, I.~Y. Park and S.-J. Sin, \emph{{Complete Einstein equations from the
  generalized First Law of Entanglement}},
  \href{http://dx.doi.org/10.1103/PhysRevD.98.026020}{\emph{Phys. Rev.} {\bf
  D98} (2018) 026020}, [\href{https://arxiv.org/abs/1709.05752}{{\tt
  1709.05752}}].

\bibitem{Faulkner:2017tkh}
T.~Faulkner, F.~M. Haehl, E.~Hijano, O.~Parrikar, C.~Rabideau and
  M.~Van~Raamsdonk, \emph{{Nonlinear Gravity from Entanglement in Conformal
  Field Theories}},
  \href{http://dx.doi.org/10.1007/JHEP08(2017)057}{\emph{JHEP} {\bf 08} (2017)
  057}, [\href{https://arxiv.org/abs/1705.03026}{{\tt 1705.03026}}].

\bibitem{Jacobson:2015hqa}
T.~Jacobson, \emph{{Entanglement Equilibrium and the Einstein Equation}},
  \href{http://dx.doi.org/10.1103/PhysRevLett.116.201101}{\emph{Phys. Rev.
  Lett.} {\bf 116} (2016) 201101},
  [\href{https://arxiv.org/abs/1505.04753}{{\tt 1505.04753}}].

\bibitem{Hollands:2012sf}
S.~Hollands and R.~M. Wald, \emph{{Stability of Black Holes and Black Branes}},
  \href{http://dx.doi.org/10.1007/s00220-012-1638-1}{\emph{Commun. Math. Phys.}
  {\bf 321} (2013) 629--680}, [\href{https://arxiv.org/abs/1201.0463}{{\tt
  1201.0463}}].

\bibitem{Lokhande:2017jik}
S.~F. Lokhande, G.~W.~J. Oling and J.~F. Pedraza, \emph{Linear response of
  entanglement entropy from holography},
  \href{http://dx.doi.org/10.1007/JHEP10(2017)104}{\emph{JHEP} {\bf 10} (2017)
  104}, [\href{https://arxiv.org/abs/1705.10324}{{\tt 1705.10324}}].

\bibitem{Jefferson:2017sdb}
R.~Jefferson and R.~C. Myers, \emph{Circuit complexity in quantum field
  theory}, \href{http://dx.doi.org/10.1007/JHEP10(2017)107}{\emph{JHEP} {\bf
  10} (2017) 107}, [\href{https://arxiv.org/abs/1707.08570}{{\tt 1707.08570}}].

\bibitem{Chapman:2017rqy}
S.~Chapman, M.~P. Heller, H.~Marrochio and F.~Pastawski, \emph{Toward a
  definition of complexity for quantum field theory states},
  \href{http://dx.doi.org/10.1103/PhysRevLett.120.121602}{\emph{Phys. Rev.
  Lett.} {\bf 120} (2018) 121602},
  [\href{https://arxiv.org/abs/1707.08582}{{\tt 1707.08582}}].

\bibitem{Camargo:2018eof}
H.~A. Camargo, P.~Caputa, D.~Das, M.~P. Heller and R.~Jefferson,
  \emph{Complexity as a novel probe of quantum quenches: universal scalings and
  purifications},
  \href{http://dx.doi.org/10.1103/PhysRevLett.122.081601}{\emph{Phys. Rev.
  Lett.} {\bf 122} (2019) 081601},
  [\href{https://arxiv.org/abs/1807.07075}{{\tt 1807.07075}}].

\bibitem{Bhattacharyya:2018bbv}
A.~Bhattacharyya, A.~Shekar and A.~Sinha, \emph{Circuit complexity in
  interacting {QFT}s and {RG} flows},
  \href{http://dx.doi.org/10.1007/JHEP10(2018)140}{\emph{JHEP} {\bf 10} (2018)
  140}, [\href{https://arxiv.org/abs/1808.03105}{{\tt 1808.03105}}].

\bibitem{Ali:2018fcz}
T.~Ali, A.~Bhattacharyya, S.~Shajidul~Haque, E.~H. Kim and N.~Moynihan,
  \emph{Time evolution of complexity: A critique of three methods},
  \href{http://dx.doi.org/10.1007/JHEP04(2019)087}{\emph{JHEP} {\bf 04} (2019)
  087}, [\href{https://arxiv.org/abs/1810.02734}{{\tt 1810.02734}}].

\bibitem{Chapman:2018hou}
S.~Chapman, J.~Eisert, L.~Hackl, M.~P. Heller, R.~Jefferson, H.~Marrochio
  et~al., \emph{Complexity and entanglement for thermofield double states},
  \href{http://dx.doi.org/10.21468/SciPostPhys.6.3.034}{\emph{SciPost Phys.}
  {\bf 6} (2019) 034}, [\href{https://arxiv.org/abs/1810.05151}{{\tt
  1810.05151}}].

\bibitem{Ge:2019mjt}
D.~Ge and G.~Policastro, \emph{{Circuit Complexity and 2D Bosonisation}},
  \href{https://arxiv.org/abs/1904.03003}{{\tt 1904.03003}}.

\bibitem{Caputa:2017urj}
P.~Caputa, N.~Kundu, M.~Miyaji, T.~Takayanagi and K.~Watanabe, \emph{Anti-de
  sitter space from optimization of path integrals in conformal field
  theories},
  \href{http://dx.doi.org/10.1103/PhysRevLett.119.071602}{\emph{Phys. Rev.
  Lett.} {\bf 119} (2017) 071602},
  [\href{https://arxiv.org/abs/1703.00456}{{\tt 1703.00456}}].

\bibitem{Caputa:2017yrh}
P.~Caputa, N.~Kundu, M.~Miyaji, T.~Takayanagi and K.~Watanabe, \emph{Liouville
  action as path-integral complexity: From continuous tensor networks to
  ads/cft}, \href{http://dx.doi.org/10.1007/JHEP11(2017)097}{\emph{JHEP} {\bf
  11} (2017) 097}, [\href{https://arxiv.org/abs/1706.07056}{{\tt 1706.07056}}].

\bibitem{Caputa:2018kdj}
P.~Caputa and J.~M. Magan, \emph{Quantum computation as gravity},
  \href{https://arxiv.org/abs/1807.04422}{{\tt 1807.04422}}.

\bibitem{Camargo:2019isp}
H.~A. Camargo, M.~P. Heller, R.~Jefferson and J.~Knaute, \emph{Path integral
  optimization as circuit complexity},
  \href{https://arxiv.org/abs/1904.02713}{{\tt 1904.02713}}.

\bibitem{dowling2008geometry}
M.~R. Dowling and M.~A. Nielsen, \emph{The geometry of quantum computation},
  {\emph{Quantum Information \& Computation} {\bf 8} (2008) 861--899}.

\bibitem{Horowitz:1998ha}
G.~T. Horowitz and R.~C. Myers, \emph{The {AdS/CFT} correspondence and a new
  positive energy conjecture for general relativity},
  \href{http://dx.doi.org/10.1103/PhysRevD.59.026005}{\emph{Phys. Rev.} {\bf
  D59} (1998) 026005}, [\href{https://arxiv.org/abs/hep-th/9808079}{{\tt
  hep-th/9808079}}].

\bibitem{Klebanov:2007ws}
I.~R. Klebanov, D.~Kutasov and A.~Murugan, \emph{{Entanglement as a probe of
  confinement}},
  \href{http://dx.doi.org/10.1016/j.nuclphysb.2007.12.017}{\emph{Nucl. Phys.}
  {\bf B796} (2008) 274--293}, [\href{https://arxiv.org/abs/0709.2140}{{\tt
  0709.2140}}].

\bibitem{Ishihara:2012jg}
M.~Ishihara, F.-L. Lin and B.~Ning, \emph{Refined holographic entanglement
  entropy for the ads solitons and ads black holes},
  \href{http://dx.doi.org/10.1016/j.nuclphysb.2013.04.003}{\emph{Nucl. Phys.}
  {\bf B872} (2013) 392--426}, [\href{https://arxiv.org/abs/1203.6153}{{\tt
  1203.6153}}].

\bibitem{Reynolds:2017jfs}
A.~P. Reynolds and S.~F. Ross, \emph{{Complexity of the AdS Soliton}},
  \href{http://dx.doi.org/10.1088/1361-6382/aab32d}{\emph{Class. Quant. Grav.}
  {\bf 35} (2018) 095006}, [\href{https://arxiv.org/abs/1712.03732}{{\tt
  1712.03732}}].

\bibitem{Momeni:2016ekm}
D.~Momeni, S.~A. Hosseini~Mansoori and R.~Myrzakulov, \emph{Holographic
  complexity in gauge/string superconductors},
  \href{http://dx.doi.org/10.1016/j.physletb.2016.03.031}{\emph{Phys. Lett.}
  {\bf B756} (2016) 354--357}, [\href{https://arxiv.org/abs/1601.03011}{{\tt
  1601.03011}}].

\bibitem{Zhang:2017nth}
S.-J. Zhang, \emph{Complexity and phase transitions in a holographic qcd
  model}, \href{http://dx.doi.org/10.1016/j.nuclphysb.2018.02.010}{\emph{Nucl.
  Phys.} {\bf B929} (2018) 243--253},
  [\href{https://arxiv.org/abs/1712.07583}{{\tt 1712.07583}}].

\bibitem{Du:2018uua}
L.-P. Du, S.-F. Wu and H.-B. Zeng, \emph{Holographic complexity of the disk
  subregion in (2+1)-dimensional gapped systems},
  \href{http://dx.doi.org/10.1103/PhysRevD.98.066005}{\emph{Phys. Rev.} {\bf
  D98} (2018) 066005}, [\href{https://arxiv.org/abs/1803.08627}{{\tt
  1803.08627}}].

\end{thebibliography}\endgroup

\end{document}